\documentclass[review]{elsarticle}

\usepackage{lineno}
\modulolinenumbers[5]

\journal{Journal of \LaTeX\ Templates}

\usepackage{graphicx,xcolor}
\usepackage[hyphenbreaks]{breakurl}
\usepackage[hyphens]{url}

\usepackage{ulem}
\usepackage[T1]{fontenc}
\usepackage[utf8]{inputenc}

\usepackage{listings}
\usepackage{adjustbox}
\usepackage{fancybox}
\usepackage{tabulary}
\usepackage{url}
\usepackage{soul}
\usepackage{amsmath}
\usepackage{amsfonts}
\usepackage{amssymb}

\usepackage{subcaption}

\lstdefinestyle{Scala}
{ 
	language=scala,
	basicstyle = \footnotesize\sffamily,
	showspaces = false,
	showstringspaces = false,
	showtabs = false,
	tabsize = 1,
	breaklines = true,
	breakatwhitespace= false,
	numbers = left,
	columns=fullflexible,
	sensitive=true,
	string=[b]",
	morestring=[d]',
	morecomment=[l]{//},
}

\newcommand*{\ldb}{\{\mskip-5mu\{}
\newcommand*{\rdb}{\}\mskip-5mu\}}
\newcommand*{\pp}{+\mskip-3mu+}

\begin{document}

\begin{frontmatter}

\title{An Abstract View of Big Data Processing Programs}
\tnotetext[thanks]{This study was financed in part by the Coordena\c c\~ ao de Aperfei\c coamento de Pessoal de N\'\i vel Superior - Brasil (CAPES) - Finance Code 001.}
\tnotetext[extention]{This paper is an extended version of~\cite{deSouzaNeto2020sbmf}.}


\author[ufrn]{Jo\~{a}o Batista de Souza Neto\corref{mycorrespondingauthor}}
\cortext[mycorrespondingauthor]{Corresponding author}
\ead{jbsneto@ppgsc.ufrn.br}

\author[ufrj]{Anamaria Martins Moreira}
\ead{anamaria@ic.ufrj.br}

\author[cnrs]{Genoveva Vargas-Solar}
\ead{genoveva.vargas-solar@liris.cnrs.fr}

\author[ufrn]{Martin A. Musicante}
\ead{mam@dimap.ufrn.br}

\address[ufrn]{Department of Informatics and Applied Mathematics (DIMAp)\\
Federal University of Rio Grande do Norte,
Natal, Brazil.}

\address[ufrj]{Institute of Computing (IC)\\
Federal University of Rio de Janeiro, Rio de Janeiro, Brazil.}

\address[cnrs]{French Council of Scientific Research (CNRS), LIRIS, Lyon, France.}

\begin{abstract}
This paper proposes a  model for specifying data flow based parallel data processing programs agnostic of target  Big Data processing frameworks. 
The paper focuses on the formal abstract specification of non-iterative and iterative programs, generalizing the strategies adopted by data flow Big Data processing frameworks.
The proposed model relies on \textit{Monoid Algebra}   and \textit{Petri Nets} to abstract Big Data processing programs in two levels:  a high level representing the program data flow and a lower level representing data transformation operations (e.g., filtering, aggregation, join). 
We extend the \textit{model for data processing programs} proposed in~\cite{deSouzaNeto2020sbmf}, to enable the use of iterative programs.
The general specification of iterative data processing programs implemented by data flow based parallel programming models is essential given the democratization of iterative and greedy Big Data analytics algorithms. Indeed, these algorithms call for revisiting parallel programming models to express iterations. The paper gives a comparative analysis of the iteration strategies proposed by Apache Spark, DryadLINQ, Apache Beam and Apache Flink. It discusses how the model achieves to generalize these strategies.

\end{abstract}


\begin{highlights}
\item This is an extended version of \textit{Modeling Big Data Processing Programs}, by João Batista de Souza Neto, Anamaria Martins Moreira, Genoveva Vargas-Solar and Martin A. Musicante. SBMF 2020. 

\item This extended version contains the following improvements, in relation to the SBMF 2020 paper:
\begin{itemize}
    \item Extension of the modeling primitives to support iterative programs.
    This is the main contribution of the extended version.
    Section~\ref{sec:ModelingIterations} present the \textit{iterate} and \textit{iterateWithCondition} primitives and their semantics in terms of Monoid Algebra.
    Iterations are represented by a loop on the Petri Net that defines the program.
    In order to have an acyclic graph to represent the program, these loops are \textit{unfolded} to build a Petri Net without cycles.
    We give an example use of these operations, by modeling a Spark programs taken from~\cite{Zaharia:2012}.
    In this example we show how the upper layer (Petri Net) of the model is unfolded to consider the new primitives.
    We conclude that the extension of the model provides more expressiveness to model both non-iterative and iterative Big Data processing algorithms. Particularly, iterative ones that come up in current analytics algorithms based on data mining, machine learning, graph analysis and artificial intelligence techniques.
    
    \item The inclusion of the description of Big Data Processing Frameworks, describing their characteristics and highlighting the strategies they propose concerning the implementation of iterative algorithms. Given the diversity of strategies adopted for addressing iteration, we discuss the importance of providing an abstract model that can represent issues related to iteration.
    The description includes, for each framework, a list of primitives they provide and their correspondence to the operations in our model (see Table~\ref{tab:systems-operations}).

    \item Besides the new technical content, the paper was revised and expanded as follows:
    The new contents include: 
    
    \textit{(i)} A better motivation and more clear explanation of the application of the model; 
    
    \textit{(ii)} Improved description of Petri Nets and Monoid Algebra. We particularly extended the description of the Monoid Algebra including the \textit{repeat} operation.
    
    \textit{(iii)} Better and more complete description of the model. In particular, we include the definition of some primitive operations that do not appear in the SBMF paper. In this sense, the extended version provides a full description of the proposed model.

    \textit{(iv)} Overall improvement of the Related Work and Conclusions. In the related work we make reference to very recent works addressing iterative algorithms in Big Data processing frameworks. Our model is complementary to these approaches since it attempts to model iteration independently of the technical characteristics of concrete target frameworks. 
\end{itemize}
\end{highlights}

\end{frontmatter}

\linenumbers

\section{Introduction}
\label{sec:intro}

The intensive processing of datasets with significant volume, variety and velocity scales, namely Big Data, calls for alternative parallel programming models adapted to the implementation of data analytics tasks and capable of exploiting the potential of those datasets.
Large-scale data processing frameworks have implemented these programming models to provide execution infrastructures giving transparent access to large scale computing and memory resources. 

Large scale data processing systems can be classified according to their purpose into general-purpose, SQL-based, graph processing, and stream processing~\cite{Bajaber2016}. 
These systems adopt different approaches to represent and process data.  
Examples of general-purpose systems are \textit{Apache Hadoop}~\cite{hadoop273}, \textit{Dryad/DryadLINQ}~\cite{Isard2007, yu2008}, \textit{Apache Flink}~\cite{katsifodimos2015apache}, \textit{Apache Beam}~\cite{beam2016} and \textit{Apache Spark}~\cite{Zaharia:2010}. 
According to the programming model adopted for processing data, general-purpose systems can be control flow-based (like Apache Hadoop) or data flow-based (like Apache Spark).
%
%
In these systems, a program is built 
from individual processing blocks.
These processing blocks implement operations that perform \textit{transformations} on the data. 
The interaction between these blocks defines the \textit{data flow} that specifies the order to perform operations. 
Datasets exchanged among the blocks are modeled by data structures such as key-value tuples or tables. 
The system infrastructure manages the parallel and distributed processing of datasets transparently. 
This facility allows developers to avoid dealing with low-level details inherent to the use of distributed and parallel environments. 

In this context, depending to the dataset properties (velocity, volume), performance expectations, and computing infrastructure characteristics (cluster, cloud, HPC nodes), it is often a critical programmer's decision to choose a well-adapted target system used for running data processing programs. 
Indeed, each hardware/software facility has its particularities concerning the infrastructure and optimizations made to run a program in a parallel and distributed way. 
This diversity suggests that systems will have different performance scores depending on their context and available resources. 
The choice between different configuration options depends on the non-functional requirements of the project, available infrastructure and even preferences of the team that develops and execute the program. 
In this context, the formulation of more abstract, platform-agnostic program descriptions could help in the design of systems that would be deployed in a variety of contexts.

In a previous paper~\cite{deSouzaNeto2020sbmf} we introduced a model for non-iterative, Big Data processing programs.
The model was proposed as an abstract view of data flow systems such as Apache Spark.
This paper extends the \textbf{\textit{model for data processing programs}} proposed in~\cite{deSouzaNeto2020sbmf}, to enable the use of iterative programs.
Our model provides an abstract representation of the main aspects of data flow-based data processing systems: 
\textit{(i)} operations applied on data (e.g., filtering, aggregation, join); \textit{(ii)} representation of programs execution through directed acyclic graphs (DAGs) where vertices represent operations and datasets, and edges represent data communication.  
In our model, a program is defined as a bipartite graph composed of transformations (i.e., operations) and datasets being processed by transformations.
When considering actual system restrictions on the predefinition of the number of iterations of any cycle, these graphs may be converted into DAGs for execution.
Our model has two levels: a high level representing the program data flow and a lower level representing data transformation operations. 

Throughout the paper, we use the name \textit{data flow} to refer to the representation of the program's data flow graph and \textit{transformations} to the operations over datasets that compose the program.
We use \textit{Petri Nets}~\cite{murata1989} to represent the data flow, and \textit{Monoid Algebra}~\cite{fegaras2017,fegaras2019} to  model transformations.
Monoid Algebra is a formal system to describe processing distributed data.
The combined use of these formalisms allows the expression of programming logic, to be implemented independently of the target Big Data processing system\footnote{Such that \textit{Apache Spark}, \textit{DryadLINQ}, \textit{Apache Beam} or \textit{Apache Flink}.}.
In this way, we provide a formal, infrastructure-independent specification of data processing programs implemented according to data flow-based programming models. 

To the extent of our knowledge, most works addressing Big Data processing programs have, so far, concentrated efforts on technical and engineering challenging aspects. 
However, few works, such as~\cite{yang2010},~\cite{Chen2017}, and~\cite{Ono2011} have worked on formal specifications that can be used to reason about their execution abstractly. 
Formal modeling parallel execution implemented by systems of the same family can be important for comparing infrastructures, defining pipelines to test parallel data processing programs, and verifying programs properties (such as correctness, completeness or concurrent access to data). 
In this work, we use the model to define mutation operators that can be instantiated for different systems.
In particular, specifications in our model have been used as an intermediate representation of programs in a mutation testing tool of Apache Spark programs~\cite{caise2020}.

Besides the introduction of iterative processing primitives, this paper extends~\cite{deSouzaNeto2020sbmf} by \textit{(i)} providing a full description of our model, including a more comprehensive use of the resources provided by Petri Nets; 
\textit{(ii)} giving a more detailed comparison of data flow-based systems, to show how they can be modeled by our proposal.

\bigskip

The remainder of the paper is organized as follows. 
Section \ref{sec:bcknd} presents the background concepts of the model, namely, Petri Nets and Monoid Algebra. 
Section \ref{sec:modeling} presents the model for formally expressing Big Data processing programs.  
Section~\ref{sec:bigdata} describes the main characteristics of data flow based  Big Data processing frameworks and discusses how our proposal can model their operations and iteration strategies. 
Section~\ref{sec:exp} describes the general lines of the way the model can be used in a concrete program testing application.
%
Section \ref{sec:relwork} introduces related work addressing approaches for generalizing control and data flow parallel programming models. 
Finally, Section \ref{sec:conc} concludes the paper and discusses future work.


\section{Background}
\label{sec:bcknd}
This section briefly presents Petri Nets and Monoid Algebra, upon which our model is built.
For a more detailed presentation, the reader can refer to~\cite{murata1989,fegaras2017}.

\paragraph{Petri Nets}\cite{Petri62} 
are a formal tool  to model and analyze the behavior of distributed, concurrent, asynchronous, and/or non-deterministic systems~\cite{murata1989}. 
A Petri Net is defined as a directed bipartite graph that contains two types of nodes: \textit{places} and \textit{transitions}. 
Places represent the system's state variables, while transitions represent the actions performed by the system. 
These two components are connected through directed edges that connect places to transitions and transitions to places. 
With these components, it is possible to represent (i) the different states of a system; (ii) the actions taken by the system to move from one state to another (transitions) (iii) and how the state changes due to actions (edges).
This modeling is done by using \textit{tokens} to decorate places of the net.
The distribution of the tokens among places indicates that the system is in a given state. 
The execution of an action (transition) takes tokens from one place to another, leading to an evolution of the system's state. 

Formally, a Petri net is a quintuple $P\!N = (P, T, F, W, M_0)$ where $P \cap T = \emptyset$, $P \cup T \neq \emptyset$ and:
\begin{eqnarray*}
P = \{p_1, p_2,  \ldots, p_m \} &{}& \mbox{ is a finite set of \textit{places},} \\
T = \{t_1, t_2,  \ldots, t_n \}  &{}& \mbox{ is a finite set of \textit{transitions},}  \\
F \subseteq (P \times T) \cup (T \times P) &{}& \mbox{ is a finite set of \textit{edges},}  \\
W: F \rightarrow \{1, 2, 3,  \ldots \}  &{}& \mbox{ is function associating positive weights to edges,} \\
M_0: P \rightarrow \{0, 1, 2, 3,  \ldots \} &{}& \mbox{ is a function defining the initial marking of a net.} 
\end{eqnarray*}
The execution of a system is defined by \textit{firing} transitions.
Firing a transition $t$ consumes $W(s, t)$ tokens from all its input places $s$, and produces $W(t, s')$ tokens to each of its output places $s'$. The transition $t$ can only be fired (it is said to be {\em enabled}) if there are at least $W(s, t)$ tokens on all its input places $s$.
The semantics of a given process is then given by the evolution of markings produced by firing enabled transitions.

\paragraph{Monoid Algebra} 
was proposed in~\cite{fegaras2017} as an algebraic formalism for data-centric distributed computing operations based on monoids and monoid homomorphisms. 
A monoid is an algebraic structure $(S, \oplus, e_\oplus)$ formed by a set $S$, an associative operation $\oplus$ in $S$ and a neutral element $e_\oplus$. 
The function $\oplus$ is usually used to identify the monoid. 
A \textit{monoid homomorphism} is a  function $H$ over two monoids, say $\otimes = (S, \otimes, e_\otimes)$ to $\oplus = (T, \oplus, e_\oplus)$, such that:

\begin{align*}
\begin{split}
H(X \otimes Y) ={}& H(X) \oplus H(Y) \quad \text{for all \,} X \text{\, and \,} Y \text{\, of \, type \,} S \\ 
H(e_\otimes) ={}& e_\oplus  
\end{split}
\end{align*}

Monoid algebra uses monoid and monoid homomorphism concepts to define operations on distributed datasets, which are represented as \textit{monoid collections}. 
One type of monoid collection is  \textit{bag}, an unordered data collection of elements of type $ \alpha $ (denoted as $ Bag[\alpha] $).
The elements of $Bag[\alpha]$ are formed by using the unit injection function $\mathbb{U}_\uplus $, which generates the unitary bag $\ldb x \rdb $ from an element $x$ ($\mathbb{U}_\uplus(x) = \ldb x \rdb $), the associative operation $ \uplus $, which unites two bags ($ \ldb x \rdb \uplus \ldb y \rdb = \ldb x, y \rdb $), and the neutral element $\ldb \rdb$, which is an empty bag. 
Another monoid collection is the one formed by lists.
It can be defined as an ordered bag.
It can be defined from the set $List[\alpha]$ containing lists of elements of type $\alpha$, and using  $\mathbb{U}_{\pp}$ as the unit injection function, the list concatenation $\pp$ as the associative operation and the empty list $[\,]$ as the neutral element of the monoid.


Monoid algebra defines distributed operations as monoid homomorphisms over monoid collections (which represent distributed datasets). 
These homomorphisms are defined to
abstractly describe the basic blocks of distributed data processing systems such as map/reduce or data flow systems.
The key idea behind monoid algebra is to use the associativity property of the monoid operations and the homomorphism between monoids to represent the processing of partitioned data and the combination of the results, independently from how data is partitioned.

\bigskip
Let us now define the most common operations used in monoid algebra.
The $ \text{\textbf{flatmap}} $ operation receives a function $ f $ of type $ \alpha \rightarrow Bag[\beta] $ and a collection $X$ of type $ Bag[\alpha] $ as input and returns a collection $ Bag[\beta] $ resulting from the union $ \uplus $ of the results of applying $f$ to each element of $X$. 
This operation captures the essence of parallel processing since $f$ can be executed in parallel on top of different data partitions in a distributed dataset.
Notice that $ \text{\textbf{flatmap}}\ f $ is a monoid homomorphism since it is a function that preserves the structure of bags. 


The operations $ \text{\textbf{groupby}} $ and $ \text{\textbf{cogroup}} $ capture the data shuffling process by representing the reorganization and grouping of data. The $ \text{\textbf{groupby}} $ operation groups the elements of $ Bag[\kappa \times \alpha] $ using the first component of type $\kappa$ as a key, resulting in a collection $ Bag[\kappa \times Bag[\alpha]] $, where the second component is a collection containing all elements of type $\alpha$ that were associated with the same key $k$ in the initial collection. 
The $\text{\textbf{cogroup}}$ operation works similarly to $\text{\textbf{groupby}}$, but it operates on two collections that have a key of the same type $\kappa$.
In this way, the result of $\text{\textbf{cogroup}}$, when applied to two collections of type $ Bag[\kappa \times \alpha] $ and $ Bag[\kappa \times \beta] $ is a collection of type $ Bag[\kappa \times Bag[\alpha \times \beta]] $. 

The $ \text{\textbf{reduce}} $ operation represents the aggregation of the elements of $ Bag[\alpha] $ into a single element of type $ \alpha $ from the application of an associative function $ f $ of type $ \alpha \rightarrow \alpha \rightarrow \alpha $. 

The operation $ \text{\textbf{orderby}} $ represents the transformation of a bag $ Bag[\kappa \times \alpha] $ into a list $ List[\kappa \times \alpha] $ ordered by the key of type $ \kappa $ which supports the total order $ \leq $. 

These operations are monoid homomorphisms, as proved in~\cite{fegaras2017}. This property makes it possible to make transparent to the model how data has been distributed when parallelizing tasks. However, they are not enough to model applications where iteration is needed. For this, Monoid algebra, as presented in~\cite{fegaras2017}, includes the $ \text{\textbf{repeat}} $ operation.

The $ \text{\textbf{repeat}} $ operation provided by Monoid Algebra is used to allow the representation of iterative algorithms~\cite{fegaras2017}, such as machine learning and graphs processing algorithms. 
The $ \text{\textbf{repeat}} $ operation receives a function $f$ of type $ Bag[\alpha] \rightarrow Bag[\alpha] $, a predicate $p$ of type $ Bag[\alpha] \rightarrow boolean $, a count number $n$, and a collection $X$ of type $ Bag[\alpha] $ as input and returns a collection of type $ Bag[\alpha] $ as output. 
The definition of $ \text{\textbf{repeat}} $ is given below~\cite{fegaras2019}:
\begin{align*} 
\text{\textbf{repeat}}(f, p, n, X) \triangleq {}& if \: n \leq 0 \: \lor \: \neg p(X) \\ 
 {}& then \: X \\
 {}& else \: \text{\textbf{repeat}}(f, p, n - 1, f(X))
\end{align*}

The $ \text{\textbf{repeat}} $ operation stops when the counter $n$ is zero or the condition $p$ in $X$ is false.  While these conditions are not met, the operation computes $f(X)$ and decrements $n$ recursively. Intuitively, in each iteration, the collection resulting from the previous iteration is processed by $f$ which produces a new collection for the next iteration (or for the output when $ \text{\textbf{repeat}} $ stops).

In addition, monoid algebra also supports the use of lambda expressions ($ \lambda x . e $), conditionals (\textbf{if-then-else}).

\bigskip
Our proposal combines the use of Petri Nets with Monoid Algebra to build abstract versions of the primitives present in Big Data processing applications.
The main goal of our approach is to have an abstract representation common to data-centric programs.
This representation may be used to compare different frameworks and as (intermediate) representation to translate, refine, or optimize programs.

\section{Modeling Big Data Processing Programs}
\label{sec:modeling}

This section introduces the proposed formal model for Big Data processing programs. 
The model is organized in two levels: \textit{data flow}, and \textit{transformations}.
Data flow in our model is defined using Petri Nets, and the semantics of the transformations applied to the data is modeled as monoid homomorphisms on datasets. 

\subsection{Data Flow}\label{sec:data_flow}
For the upper level of our two-level modelization, we define a graph representing the data flow of a data processing program.
We rely on the data flow graph model presented in~\cite{kavi1986}, which was formalized using Petri Nets~\cite{murata1989}. 

A  program $P$ is defined as a bipartite directed graph where places stand for the distributed datasets ($D$) of the program, and transitions stand for its transformations ($T$). 
Datasets and transformations are connected by edges ($E$):
\begin{center}
$ P = \langle D \cup T, E \rangle $
\end{center}

This graph can be seen as a Petri Net, as defined in Section~\ref{sec:bcknd}. Datasets correspond to the places of the net and transformations correspond to the net transitions. The initial marking ($M_0$) of the Petri Net represents the availability of the input datasets for the computation to begin.
There will be as many tokens in an input dataset as the number of uses of this dataset in the program.
The weight function $W$ is defined as $1$ for every edge leaving a place and as $k$, for every edge arriving at a place, where $k$ is the number of times the exact same dataset is used in the program. 
That is, for each edge $(t, d)\in E,\ W(t, d) = |O(d)|$ and for each edge  $(t, d)\in E,\  W(d, t) = 1$, where $O(d)$ represents the set of transformations that receive $d$ as input, \textit{i.e.}, the number o edges coming out of $d$.

\begin{figure}[!htbp]
	\centering
	\begin{subfigure}[t]{0.5\textwidth}
		\centering
		\includegraphics[width=.6\textwidth]{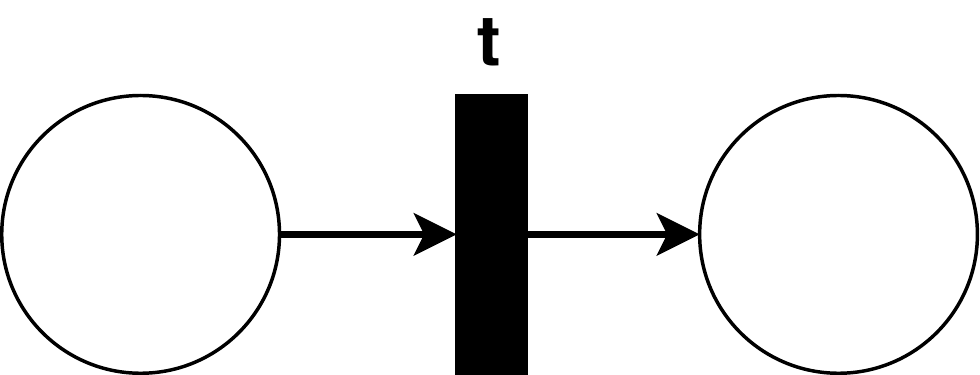}
		\caption{Unary Transformation.}
		\label{subfig:unary-transformation}
	\end{subfigure}%
 	\hfill
	\begin{subfigure}[t]{0.5\textwidth}
		\centering
		\includegraphics[width=.6\textwidth]{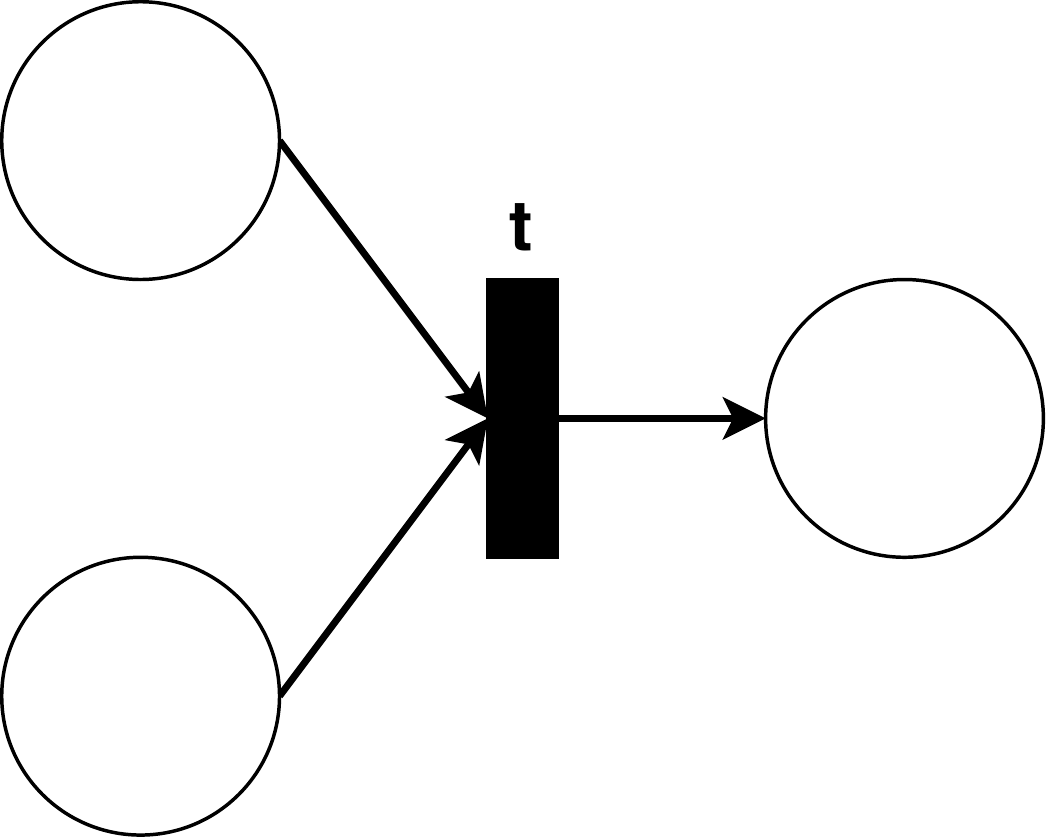}
		\caption{Binary Transformation.}
		\label{subfig:binary-transformation}
	\end{subfigure}
	\caption{Types of transformations in the data flow.}
	\label{fig:dataflow-transformations}%
\end{figure}

For the purpose of constructing the data flow model, the available transformations on the modeled frameworks fall into two categories: basic transformations (without cycles) and iterative transformations. We first present the more common case of acyclic programs. The extension of our model to deal with iterations is presented in Section~\ref{sec:ModelingIterations}. 
All basic transformations in our model can have their data flow modeled by either a single transition with one input and one output edges  (see Figure~\ref{subfig:unary-transformation}) or a single transition with two input and one output edges (see Figure~\ref{subfig:binary-transformation}). 
We call \textit{unary transformations} those that receive only one dataset as input and \textit{binary transformations} those that receive two datasets as input.
To construct the complete graph (actually a DAG), the transitions are to be sequenced by matching the corresponding input and output datasets.


To illustrate the model, let us consider the Spark program shown in Figure~\ref{fig:union-logs-example}. 
This program receives as input two datasets (RDDs) containing log messages (line 1).
It makes the union of these two datasets (line 2), removes duplicate logs (line 3), and ends by filtering headers, removing logs that match a specific pattern (line 4).
The program ends by returning the filtered RDD (line 5). 

\begin{figure}
\centering
\begin{lstlisting}[style=Scala, linewidth=.7\textwidth]
def unionLogsExample(firstLogs: RDD[String], secondLogs: RDD[String]): RDD[String] = {
	val aggregatedLogLines = firstLogs.union(secondLogs)
	val uniqueLogLines = aggregatedLogLines.distinct()
	val cleanLogLines = uniqueLogLines.filter((line: String) => !(line.startsWith("host") && line.contains("bytes")))
	return cleanLogLines
}
\end{lstlisting}     
\caption{Sample log union program in Spark.}
\label{fig:union-logs-example}
\end{figure}


In this program, we can identify five RDDs, that will be referred to using short names for conciseness. So, 
$
D = \{d_1, d_2, d_3, d_4, d_5 \} 
$, 
where $d_1 =$ \texttt{firstLogs}, $d_2 =$ \texttt{secondLogs}, $d_3 =$ \texttt{aggregatedLogLines}, $d_4 =$ \texttt{uniqueLogLines}, and $ d_5 =$ \texttt{cleanLogLines}. 
For simplicity, each RDD in the code was given a unique name. It makes it easier to reference them in the text. However, the model considers that each RDD is uniquely identified, independently of the concrete name given to it in the code. 

We can also identify the application of three transformations in $P$; thus the set $T$ in our example is defined as
$
T = \{t_1, t_2, t_3\} 
$,
where $t_1 = \texttt{union}(d_1, d_2) $, $t_2 = \texttt{distinct}(d_3) $, and $t_3 = \texttt{filter}(\texttt{(line: String) => }$ $\texttt{!(line.startsWith}$ $\texttt{(``host'')}$ $\texttt{ \&\& line.contains(``bytes''))}, d_4)$.

Each transformation in $T$ receives one or two datasets belonging to $D$ as input and produces a dataset also in $D$ as output. 
Besides, the sets $D$ and $T$ are disjoint and finite.

Edges connect datasets with transformations. 
An edge may either be a pair in $D \times T$, representing the input dataset of a transformation, or it can be a pair in $T \times D$, representing the output dataset of a transformation. 
In this way, the set of edges of $P$ is defined as $ E \subseteq (D \times T) \cup (T \times D)$.

The set $E$ in our example program is, then:
\begin{center}
	$ E = \{(d_1, t_1), (d_2, t_1), (t_1, d_3), (d_3, t_2), (t_2, d_4), (d_4, t_3), (t_3, d_5)\} $
\end{center}

\begin{figure}[!htbp]
	\centering
	\includegraphics[width=.5\textwidth]{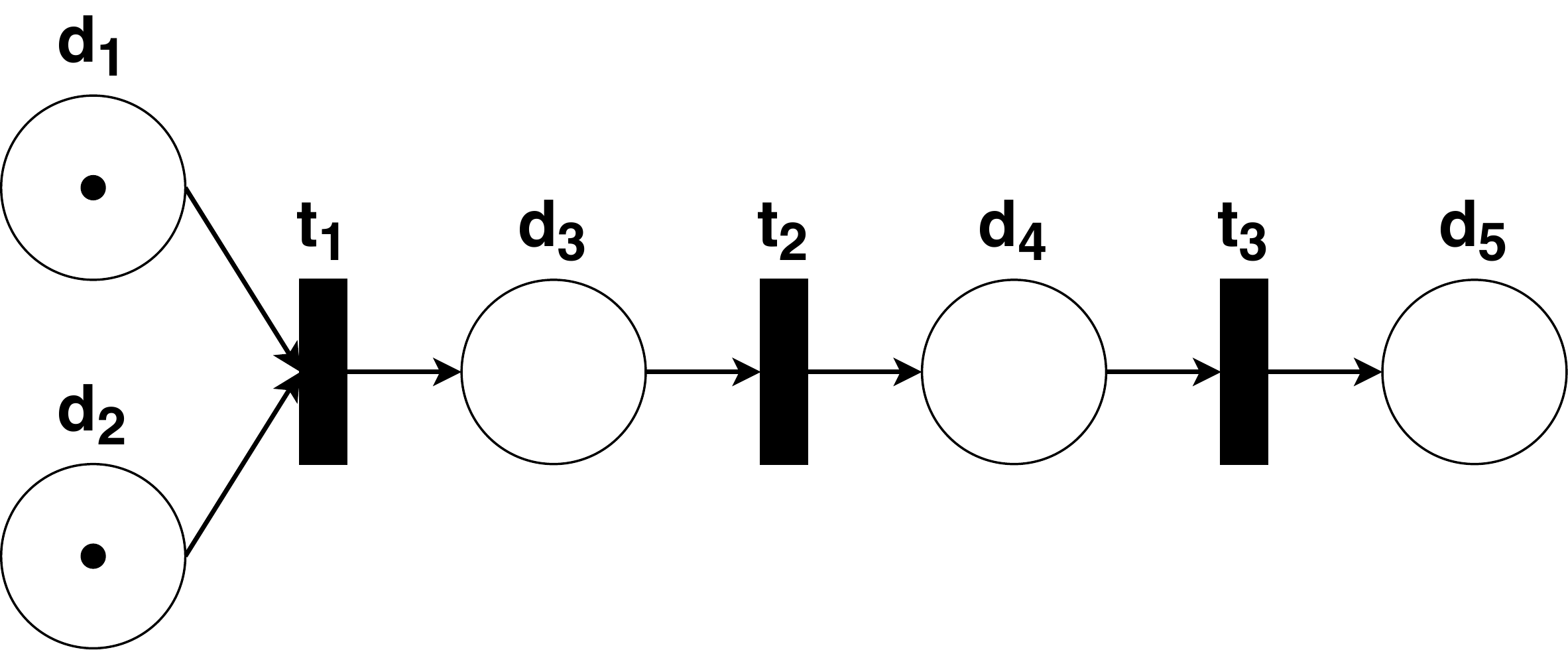}
	\caption{Data flow representation of the program in Figure~\ref{fig:union-logs-example}.}
	\label{fig:union-logs-dataflow}
\end{figure}

Using these sets, we can define a graph representing the Spark program in Figure~\ref{fig:union-logs-example}.
This graph is depicted in Figure~\ref{fig:union-logs-dataflow}. 
The distributed datasets in $D$ are represented as circle nodes, and the transformations in $T$ are represented as thick bar nodes of the graph, as it is usual in representing Petri Nets. 
The edges are represented by arrows that connect the datasets and transformations. 
The token marking in $d_1$ and $d_2$ indicate that the program is ready to be executed (initial marking).
For simplicity, we only indicate the weight of edges of the Petri Net when they are different from 1.

\subsection{Data Sets and Transformations}
The data flow model defined above represents (i) the datasets and transformations of a program $P$; (ii) the order in which transformations are processed when the program $P$ is executed. These representations are abstract from their actual contents or semantics. 

To define the contents of datasets in $D$ and the semantics of transformations in $T$, we make use of \textit{Monoid Algebra} ~\cite{fegaras2017,fegaras2019}. 
Datasets are represented as monoid collections, and transformations are defined as operations supported by monoid algebra. These representations are detailed in the following.

\subsubsection{Distributed Datasets}

A distributed dataset in $ D $ can either be represented by a \textit{bag} ($ Bag [\alpha] $) or a \textit{list} ($ List [\alpha] $). 
Both structures represent collections of distributed data ~\cite{fegaras2019}, capturing the essence of the concepts of \textit{RDD} in Apache Spark, \textit{PCollection} in Apache Beam, \textit{DataSet} in Apache Flink and \textit{DryadTable} in DryadLINQ. These structures provide an abstraction of the actual distributed data in a cluster in the form of a simple collection of items.


We define most of the transformations of our model in terms of bags. We consider lists only for transformations implementing sorts, which are the only ones in which the order of the elements in the dataset is relevant.


In monoid algebra, bags and lists can either represent distributed or local collections. 
Monoid homomorphisms treat these two kinds of collections in a unified way~\cite{fegaras2019}.
In this way, we will not distinguish between distributed and local collections when defining our transformations.


\subsubsection{Transformations}

In our model, transformations on datasets take one or two datasets as input and produce one dataset as an output. 
Transformations may also receive other types of parameters such as functions, which represent data processing operations defined by the developer and literals such as boolean constants. 
A transformation $t$ in the transformation set $T$ of a program $P$ is characterized by (i) the operation it implements, (ii) the types of its input and output datasets, (iii) and its input parameters.

We define the transformations of our model in terms of the operations of monoid algebra defined in Section~\ref{sec:bcknd}. 
We group transformations into categories according to the types of operations that we identified in the data processing systems that we studied.

\paragraph{Mapping Transformations}
transform values of an input dataset into values of an output dataset by applying a mapping function. 
Our model provides two mapping transformations: \textit{flatMap} and \textit{map}. 
Both transformations apply a given function $f$ to every element of the input dataset to generate the output dataset, the only difference being the requirements on the type of $f$ and its relation with the type of the generated dataset. 
Given an input dataset of type $Bag[\alpha]$, the \textit{map} transformation accepts any $f: \alpha \rightarrow \beta$ and generates an output dataset of type $Bag [\beta]$, while the \textit{flatMap} transformation requires $f: \alpha \rightarrow Bag[\beta]$ to produce a dataset of type $Bag [\beta]$ as output.

The definition of \textit{flatMap} in our model is just the monoid algebra operation defined in Section~\ref{sec:bcknd}:
\begin{align*}
\begin{split}
\textit{flatMap} ::{}& (\alpha \rightarrow Bag[\beta]) \rightarrow Bag[\alpha] \rightarrow Bag[\beta] \\
\textit{flatMap}(f, D) ={}& \text{\textbf{flatmap}}(f, D) 
\end{split}
\end{align*}

The  \textit{map} transformation derives data of type $Bag [\beta]$ when given a function $ f: \alpha \rightarrow \beta$. For that to be modeled with the \textbf{flatmap} from monoid algebra, we create a lambda expression that receives an element $x$ from the input dataset and results in a $ Bag [\beta]$ collection containing only the result of applying $f$ to $x$ ($ \lambda x. \ldb f (x) \rdb $). Thus, \textit {map} is defined as:
\begin{align*}
\begin{split}
map ::{}& (\alpha \rightarrow \beta) \rightarrow Bag[\alpha] \rightarrow Bag[\beta] \\
map(f, D) ={}& \text{\textbf{flatmap}}(\lambda x . \ldb f(x) \rdb , D) 
\end{split}
\end{align*}

\paragraph{Filter Transformation}
uses a boolean function to determine whether a data item should be mapped to the output dataset. 
As in the case of \textit{map}, we use a lambda expression to build a singleton bag: 
\begin{align*}
\begin{split}
filter ::{}& (\alpha \rightarrow boolean ) \rightarrow Bag[\alpha] \rightarrow Bag[\alpha] \\
filter(p, D) ={}& \text{\textbf{flatmap}}(\lambda x . \text{ \textbf{if} } p(x) \text{ \textbf{then} } \ldb x \rdb  \text{ \textbf{else} } \ldb  \rdb , D)
\end{split}
\end{align*}

For each element $x$ of the input dataset $D$, the \textit{filter} transformation checks the condition $p(x)$. 
It forms the singleton bag $ \ldb x \rdb $ or the empty bag ($ \ldb \rdb $), depending on the result of that test.
This lambda expression is then applied to the input dataset using the $ \text {\textbf {flatmap}} $ operation.

For instance, consider the boolean function $ p(x) = x \geq 3 $ and a bag $ D = \ldb 1,2,3,4,5 \rdb  $. 
then, $ filter(p, D) = \ldb 3, 4, 5 \rdb  $.

\paragraph{Grouping Transformations}
group the elements of a dataset with respect to a key. 
We define two grouping transformations in our model: \textit{groupByKey} and \textit{groupBy}.
The \textit{groupByKey} transformation is defined as the \textbf{groupby} operation of Monoid Algebra. 
It maps a key-value dataset into a dataset associating each key to a bag.
Our \textit{groupBy} transformation uses a function $ k $ to map elements of the collection to a key \textit{before} grouping the elements with respect to that key:
\begin{align*}
\begin{split}
groupBy ::{}& (\alpha \rightarrow \kappa ) \rightarrow Bag[\alpha] \rightarrow Bag[\kappa \times Bag[\alpha]] \\
groupBy(k, D) ={}& \text{\textbf{groupby}}(\text{\textbf{flatmap}}(\lambda x . \ldb (k(x), x) \rdb , D))  \\[5pt]
groupByKey ::{}& Bag[\kappa \times \alpha] \rightarrow Bag[\kappa \times Bag[\alpha]] \\
groupByKey(D) ={}& \text{\textbf{groupby}}(D)
\end{split}
\end{align*}
For example, let us consider the identity function to define each key, and the datasets $ D_1 = \ldb 1, 2, 3, 2, 3 , 3 \rdb $, and $ D_2 = \ldb (1, a), (2, b), (3, c), (1, e), (2, f) \rdb $. 
Applying \textit{groupBy} and \textit{groupByKey} to these sets results in:
\begin{align*}
\begin{split}
groupBy(\lambda k. k, D_1) ={}& \ldb (1,\ldb 1 \rdb ), (2, \ldb 2, 2 \rdb ), (3, \ldb 3, 3, 3 \rdb ) \rdb  \\
groupByKey(D_2) ={}& \ldb (1, \ldb a, e \rdb ), (2, \ldb b, f \rdb ), (3, \ldb c \rdb ) \rdb 
\end{split}
\end{align*}

\paragraph{Set-like Transformations}
correspond to binary mathematical operations in distributed collections such as those defined in set theory.
They operate on two datasets of the same type and result in a new dataset of the same type. 
The definition of these transformations is based on the definitions in~\cite{fegaras2019}. 


The \textit {union} transformation represents the union of elements from two datasets into a single dataset. This operation is represented in a simple way using the \textit {bags} union operator ($ \uplus $):
\begin{align*}
\begin{split}
union ::{}&  Bag[\alpha] \rightarrow Bag[\alpha] \rightarrow Bag[\alpha] \\
union(D_x, D_y) ={}& D_x \uplus D_y
\end{split}
\end{align*}

We also define the \textit{intersection} and \textit{subtract} transformations. To define these transformations, we first define auxiliary operations \textit {some} and \textit {all} that represent the existential ($ \exists $) and universal ($ \forall $) quantifiers, respectively. These operations receive a predicate function $p$ and reduce the dataset to a logical value:
\begin{align*}
\begin{split}
some ::{}& (\alpha \rightarrow boolean ) \rightarrow Bag[\alpha] \rightarrow boolean \\
some(p, D) ={}& \text{\textbf{reduce}}(\lor, t_1(p, D)) \\
t_1(p, D) ={}& \text{\textbf{flatmap}}(\lambda x . \ldb p(x) \rdb , D) \\ 
all ::{}& (\alpha \rightarrow boolean ) \rightarrow Bag[\alpha] \rightarrow boolean \\
all(p, D) ={}& \text{\textbf{reduce}}(\land, t_1(p, D)) \\
t_1(p, D) ={}& \text{\textbf{flatmap}}(\lambda x . \ldb p(x) \rdb , D)
\end{split}
\end{align*}

Using \textit {some} and \textit {all}, we can define the transformations \textit {intersection} and \textit {subtract} as follows:
\begin{align*}
intersection ::{}& Bag[\alpha] \rightarrow Bag[\alpha] \rightarrow Bag[\alpha] \\
\begin{split}
intersection(D_x, D_y) ={}& \text{\textbf{flatmap}}(\lambda x . \text{ \textbf{if} } some(\lambda y . x = y, D_y) \\
& \qquad \qquad \qquad \qquad \qquad \text{ \textbf{then} } \ldb  x  \rdb  \text{ \textbf{else} } \ldb  \rdb  ), D_x) 
\end{split}
\\ 
\begin{split}
subtract ::{}&  Bag[\alpha] \rightarrow Bag[\alpha] \rightarrow Bag[\alpha] \\
subtract(D_x, D_y) ={}& \text{\textbf{flatmap}}(\lambda x . \text{ \textbf{if} } all(\lambda y . x \neq y, D_y) \\
& \qquad \qquad \qquad \qquad \qquad  \text{ \textbf{then} } \ldb  x  \rdb  \text{ \textbf{else} } \ldb  \rdb  ), D_x)
\end{split}
\end{align*}

The \textit{intersection} of bags $D_x$ and $D_y$ selects all elements of $D_x$ appearing at least once in $D_y$.
Subtracting $D_y$ from $D_x$ selects all the elements of $D_x$ that differ from every element of $D_y$.

Unlike the union operation in mathematical sets, the \textit{union} transformation defined in our model maintains repeated elements from the two input datasets. To allow the removal of these repeated elements, we define the \textit{distinct} transformation. To define \textit{distinct}, we first map each element of the dataset to a key/value tuple containing the element itself as a key. After, we group this key/value dataset, which will result in a dataset in which the group is the repeated key itself. Last, we map the key/value elements only to the key, resulting in a dataset with no repetitions. The \textit{distinct} transformation is defined as follows:
\begin{align*}
\begin{split}
distinct ::{}& Bag[\alpha] \rightarrow Bag[\alpha] \\
distinct(D) ={}& \text{\textbf{flatmap}}(\lambda (k, g) . \ldb k \rdb , t_2(D)) \\
t_1(D) ={}& \text{\textbf{flatmap}}(\lambda x . \ldb (x, x) \rdb , D) \\
t_2(D) ={}& \text{\textbf{groupby}}(t_1(D)) 
\end{split}
\end{align*}

\paragraph{Aggregation Transformations}
collapses elements of a dataset into a single element. 
The most common aggregations apply binary operations on the elements of a dataset to generate a single element, resulting in a single value or on groups of values associated with a key. 
We represent these aggregations with the transformations \textit {reduce}, which operates on the whole set, and \textit {reduceByKey}, which operates on values grouped by key. 
The \textit {reduce} transformation has the same behavior as the $ \text {\textbf {reduce}} $ operation of monoid algebra.  
The definition of \textit {reduceByKey} is also defined in terms of $ \text {\textbf {reduce}} $, but since its result  is the aggregation of elements associated with each key rather than the aggregation of all elements of the set, we first need to group the elements of the dataset by their keys:
\begin{align*}
\begin{split}
reduce ::{}& (\alpha \rightarrow \alpha \rightarrow \alpha) \rightarrow Bag[ \alpha] \rightarrow \alpha \\
reduce(f, D) ={}&  \text{\textbf{reduce}}(f, D) \\[10pt]
reduceByKey ::{}& (\alpha \rightarrow \alpha \rightarrow \alpha) \rightarrow Bag[\kappa \times \alpha] \rightarrow Bag[\kappa \times \alpha] \\
reduceByKey(f, D) ={}& \text{\textbf{flatmap}}(\lambda (k, g) . \ldb (k, \text{\textbf{reduce}}(f, g)) \rdb , \text{\textbf{groupby}}(D))
\end{split}
\end{align*}


\paragraph{Join Transformations}
implement relational join operations between two datasets. 
We define four join operations, which correspond to well-known operations in relational databases: \textit{innerJoin}, \textit{leftOuterJoin}, \textit{rightOuterJoin}, and \textit{fullOuterJoin}.
The \textit{innerJoin} operation combines the elements of two datasets based on a join-predicate expressed as a relationship, such as the same key.
\textit{LeftOuterJoin} and \textit{rightOuterJoin} 
combine the elements of two sets like an \textit{innerJoin} adding to the result all values in the left (right) set that do not match to the right (left) set.
The \textit{fullOuterJoin} of two sets forms a new relation containing all the information present in both sets. 

See below  the definition of the \textit{innerJoin} transformation, which was based on the definition presented in~\cite{chlyah2019}:
\begin{align*}
\begin{split}
innerJoin ::{}& Bag[\kappa \times \alpha] \rightarrow Bag[\kappa \times \beta] \rightarrow Bag[\kappa \times (\alpha \times \beta)] \\
innerJoin(D_x, D_y) ={}& \text{\textbf{flatmap}}(\lambda (k, (d_x, d_y)) . t_2(k, d_x, d_y), t_1(D_x, D_y)) \\
t_1(D_x, D_y) ={}& \text{\textbf{cogroup}}(D_x, D_y) \\
t_2(k, d_x, d_y) ={}& \text{\textbf{flatmap}}(\lambda x . t_3(k, x, d_y), d_x) \\
t_3(k, x, d_y) ={}& \text{\textbf{flatmap}}(\lambda y . \ldb (k, (x,y)) \rdb , d_y)
\end{split}
\end{align*}

The definition of the other joins follows a similar logic, but conditionals are included to verify the different relationships. In cases where one side does not have pairs with a certain key, the result of the join is an empty bag on that side and the element that has the key on the other side. The definitions of \textit{leftOuterJoin}, \textit{rightOuterJoin}, and \textit{fullOuterJoin} are as follows:
\begin{align*}
\begin{split}
leftOuterJoin ::{}& Bag[\kappa \times \alpha] \rightarrow Bag[\kappa \times \beta] \rightarrow Bag[\kappa \times (\alpha \times Bag[\beta])] \\
leftOuterJoin(D_x, D_y) ={}& \text{\textbf{flatmap}}(\lambda (k, (d_x, d_y)) . t_2(k, d_x, d_y), t_1(D_x, D_y)) \\
t_1(D_x, D_y) ={}& \text{\textbf{cogroup}}(D_x, D_y) \\
t_2(k, d_x, d_y) ={}& \text{\textbf{if} } d_y = \ldb  \rdb  \text{ \textbf{then} } t_3(k, d_x)  \text{ \textbf{else} } t_4(k, d_x, d_y) \\
t_3(k, d_x) ={}& \text{\textbf{flatmap}}(\lambda x . \ldb (k, (x, \ldb  \rdb )) \rdb , d_x) \\
t_4(k, d_x, d_y) ={}& \text{\textbf{flatmap}}(\lambda x . t_5(k, x, d_y), d_x) \\
t_5(k, x, d_y) ={}& \text{\textbf{flatmap}}(\lambda y . \ldb (k, (x,\ldb y \rdb )) \rdb , d_y)
\end{split}
\end{align*}

\begin{align*}
\begin{split}
rightOuterJoin ::{}& Bag[\kappa \times \alpha] \rightarrow Bag[\kappa \times \beta] \rightarrow Bag[\kappa \times (Bag[\alpha] \times \beta)] \\
rightOuterJoin(D_x, D_y) ={}& \text{\textbf{flatmap}}(\lambda (k, (d_x, d_y)) . t_2(k, d_x, d_y), t_1(D_x, D_y)) \\
t_1(D_x, D_y) ={}& \text{\textbf{cogroup}}(D_x, D_y) \\
t_2(k, d_x, d_y) ={}& \text{\textbf{if} } d_x = \ldb  \rdb  \text{ \textbf{then} } t_3(k, d_y)  \text{ \textbf{else} } t_4(k, d_x, d_y) \\
t_3(k, d_y) ={}& \text{\textbf{flatmap}}(\lambda y . \ldb (k, (\ldb  \rdb , y)) \rdb , d_y) \\
t_4(k, d_x, d_y) ={}& \text{\textbf{flatmap}}(\lambda x . t_5(k, x, d_y), d_x) \\
t_5(k, x, d_y) ={}& \text{\textbf{flatmap}}(\lambda y . \ldb (k, (\ldb x \rdb , y)) \rdb , d_y)
\end{split}
\end{align*}

\begin{align*}
\begin{split}
fullOuterJoin ::{}& Bag[\kappa \times \alpha] \rightarrow Bag[\kappa \times \beta] \rightarrow Bag[\kappa \times (Bag[\alpha] \times Bag[\beta])] \\
fullOuterJoin(D_x, D_y) ={}& \text{\textbf{flatmap}}(\lambda (k, (d_x, d_y)) . t_2(k, d_x, d_y), t_1(D_x, D_y)) \\
t_1(D_x, D_y) ={}& \text{\textbf{cogroup}}(D_x, D_y) \\
t_2(k, d_x, d_y) ={}& \text{\textbf{if} } d_x \neq \ldb  \rdb  \land d_y = \ldb  \rdb  \text{ \textbf{then} } t_3(k, d_x) \text{ \textbf{else} } t_4(k, d_x, d_y) \\
t_3(k, d_x) ={}& \text{\textbf{flatmap}}(\lambda x . \ldb (k, (\ldb x \rdb , \ldb  \rdb )) \rdb , d_x) \\
t_4(k, d_x, d_y) ={}& \text{\textbf{if} } d_x = \ldb  \rdb  \land d_y \neq \ldb  \rdb  \text{ \textbf{then} } t_5(k, d_y) \text{ \textbf{else} } t_6(k, d_x, d_y) \\
t_5(k, d_y) ={}& \text{\textbf{flatmap}}(\lambda y . \ldb (k, (\ldb  \rdb , \ldb y \rdb )) \rdb , d_y) \\
t_6(k, d_x, d_y) ={}& \text{\textbf{flatmap}}(\lambda x . t_7(k, x, d_y), d_x) \\
t_7(k, x, d_y) ={}& \text{\textbf{flatmap}}(\lambda y . \ldb (k, (\ldb x \rdb ,\ldb y \rdb )) \rdb , d_y) 
\end{split}
\end{align*}



\paragraph{Sorting Transformations} 
add the notion of \textit{order} to a bag.
In practical terms, these operations receive a bag and form a list, ordered according to some criteria.
Sort transformations are defined in terms of the $ \text {\textbf {orderby}} $ operation of monoid algebra, which transforms a $ Bag [\kappa \times \alpha] $  into a $ List [\kappa \times \alpha] $ ordered by the key of type $ \kappa $ that supports the total order $ \leq $ (we will also use the $ inv $ function, which reverses the total order of a list, thus using $\geq$ instead of $ \leq $). 
We define two transformations, the \textit{orderBy} transformation that sorts a dataset of type $\alpha$, and the \textit{orderByKey} transformation that sorts a key/value dataset by the key.
The definitions of our sorting transformations are as follows:
\begin{align*}
\begin{split}
orderBy ::{}& boolean \rightarrow Bag[\alpha] \rightarrow List[\alpha] \\
orderBy(desc, D) ={}& \text{\textbf{flatmap}}(\lambda (k, v) . [k], \text{\textbf{orderby}}(t_1(desc, D))) \\
t_1(desc, D) ={}& \text{\textbf{if} } desc \text{ \textbf{then} } t_2(D) \text{ \textbf{else} } t_3(D) \\
t_2(D) ={}& \textbf{\text{\textbf{flatmap}}}(\lambda x . \ldb (inv(x), x) \rdb , D) \\
t_3(D) ={}& \text{\textbf{flatmap}}(\lambda x . \ldb (x, x) \rdb , D) \\
  {}& \\
orderByKey ::{}& boolean \rightarrow Bag[\kappa \times \alpha] \rightarrow List[\kappa \times \alpha] \\
orderByKey(desc, D) ={}& \text{\textbf{orderby}}(t_1(desc, D)) \\
t_1(desc, D) ={}& \text{\textbf{if} } desc \text{ \textbf{then} } t_2(D) \text{ \textbf{else} } D \\
t_2(D) ={}& \text{\textbf{flatmap}}(\lambda (k, x) . \ldb (inv(k), x) \rdb , D)
\end{split}
\end{align*}

The boolean value used as first parameter defines if the direct order $\leq$ or its inverse is used.

To exemplify the use of sorting transformations let us consider  $ D_1 = \ldb 1,3,2,5,4 \rdb  $ and  $ D_2 = \ldb (1, a), (3, c), (2, a), (5,e), (4, d) \rdb $. 
Then:
\begin{align*}
orderBy(false, D_1) ={}& [1, 2, 3, 4, 5] \\
orderBy(true, D_1) ={}& [5, 4, 3, 2, 1] \\
orderByKey(false, D_2) ={}& [(1, a), (2, b), (3, c), (4, d), (5, e)] \\
orderByKey(true, D_2) ={}& [(5, e), (4, d), (3, c), (2, b), (1, a)] 
\end{align*}


\subsection{Modeling Iterative Programs}
\label{sec:ModelingIterations}
Iterative algorithms apply an operation repeatedly until a predetermined number of iterations or given conditions are reached. Common iterative algorithms are machine learning algorithms, such as \textit{Logistic Regression}~\cite{hastie2009elements}, and graph analysis algorithms, such as \textit{PageRank}~\cite{brin1998anatomy}, which perform iterative optimizations and calculations.

Big Data processing systems like Apache Spark, Apache Flink, Apache Beam and Dryad/DryadLINQ represent their programs as DAGs (Directed Acyclic Graphs). These systems apply a lazy evaluation strategy to execute programs. Thus, the programs are first defined, then they are translated into an optimized DAG representing the execution plan, and, finally, they are sent to run in parallel. Due to this characteristic, iterative programs, characterized by cycles, must be translated into a DAG. Therefore, the operations executed iteratively in the program must be repeated $n$ times in the DAG, where $n$ is the number of iterations performed by the program.

In the systems Apache Spark, Apache Beam and Dryad/DryadLINQ,  iterative programs are defined with the aid of loop statements  (such as \textit{for} and \textit{while}) of the underlying programming language to control iterations. Apache Flink, on the other hand, has a native operation (\textit{iterate}) for that, where iterative operations must be encapsulated in a step function that is performed a predetermined number of times or until a specific condition, given by a convergence function, is reached.

Our model relies on the Apache Flink approach to represent the data flow of iterative programs. We define the transformations to be executed iteratively, encapsulating them in a step function that will be repeated as many times as specified in the program. The input and output of the step function must be datasets of the same type so that the output of an iteration is an input for the next one. 

\begin{figure}
\centering
\includegraphics[width=.7\textwidth]{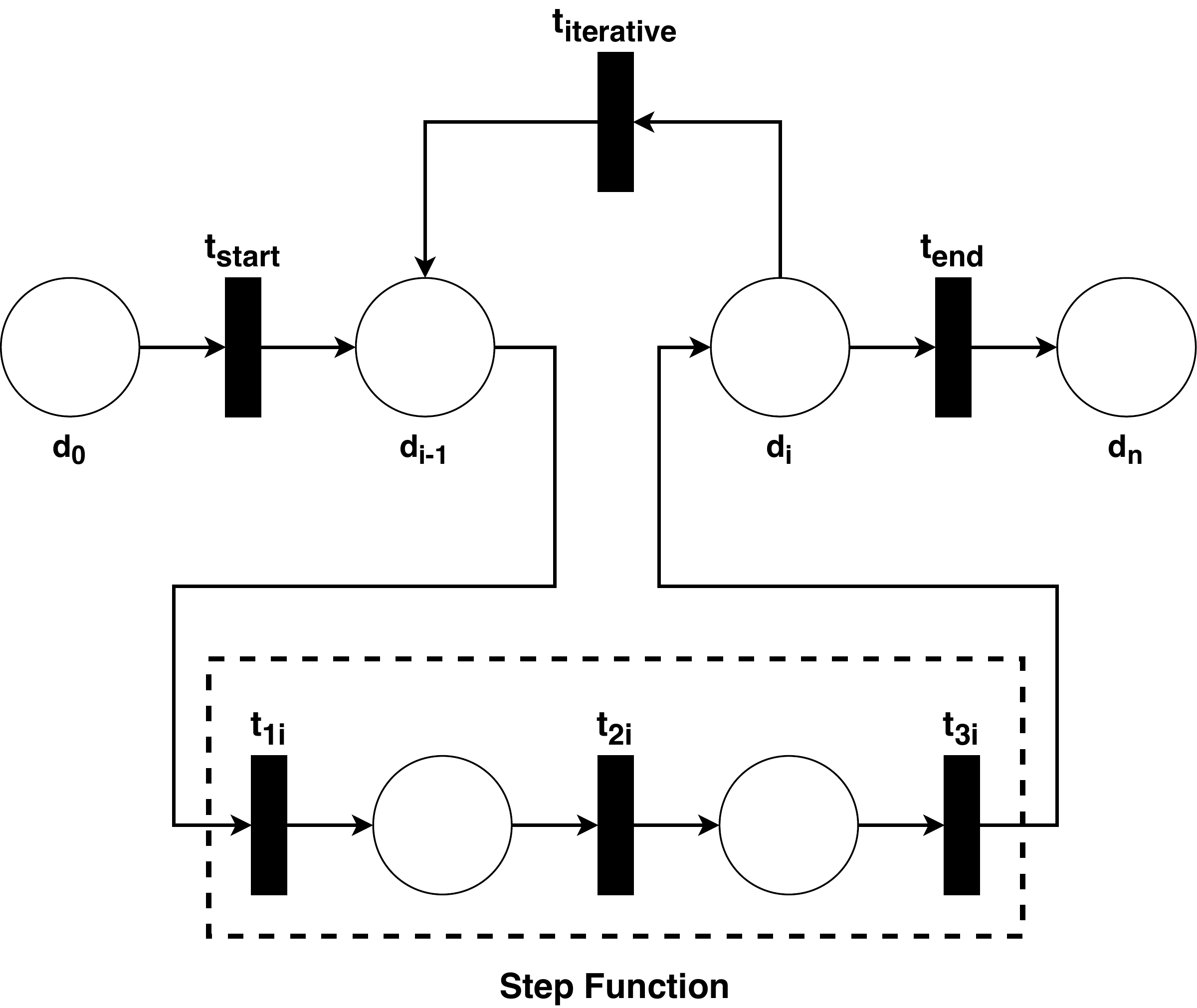}
\caption{Iterative data flow.}
\label{fig:iterate-petri}
\end{figure}

\paragraph{Iterative Data Flow}
to represent the data flow of an iterative program, we use auxiliary transitions to represent the beginning of the iterations ($t_{start}$), the repetition of the step function through a cycle in the graph ($t_{iterative}$) and the end of the iterations ($t_{end}$). In practice, these transitions are identity transformations since they do not make changes to the data, but only control the iterations. We assume that the iteration starts with an input dataset $d_0$ and that the step function will be executed $n$ times, resulting in the dataset $d_n$ as output. In the data flow model, we abstract the control of the number of iterations. Thus, the number of iterations in the data flow model is non-deterministic. We delegate this control for a specific transformation that will be presented later. Figure~\ref{fig:iterate-petri} shows how the data flow of an iterative program is represented in our model. We highlight the step function with dashed lines to represent the part repeated in each iteration.


Each iteration data flow is represented by such a sub-net and, to construct the complete Petri Net for a program, it must be composed with the other transformations as was the case with acyclic transformations. The place corresponding to its initial dataset ($d_0$) is the output from some previous transition and the place corresponding to its final dataset ($d_n$) is the input for a transition in its sequence or a final (output) place.

\begin{figure}
\centering
\includegraphics[width=.7\textwidth]{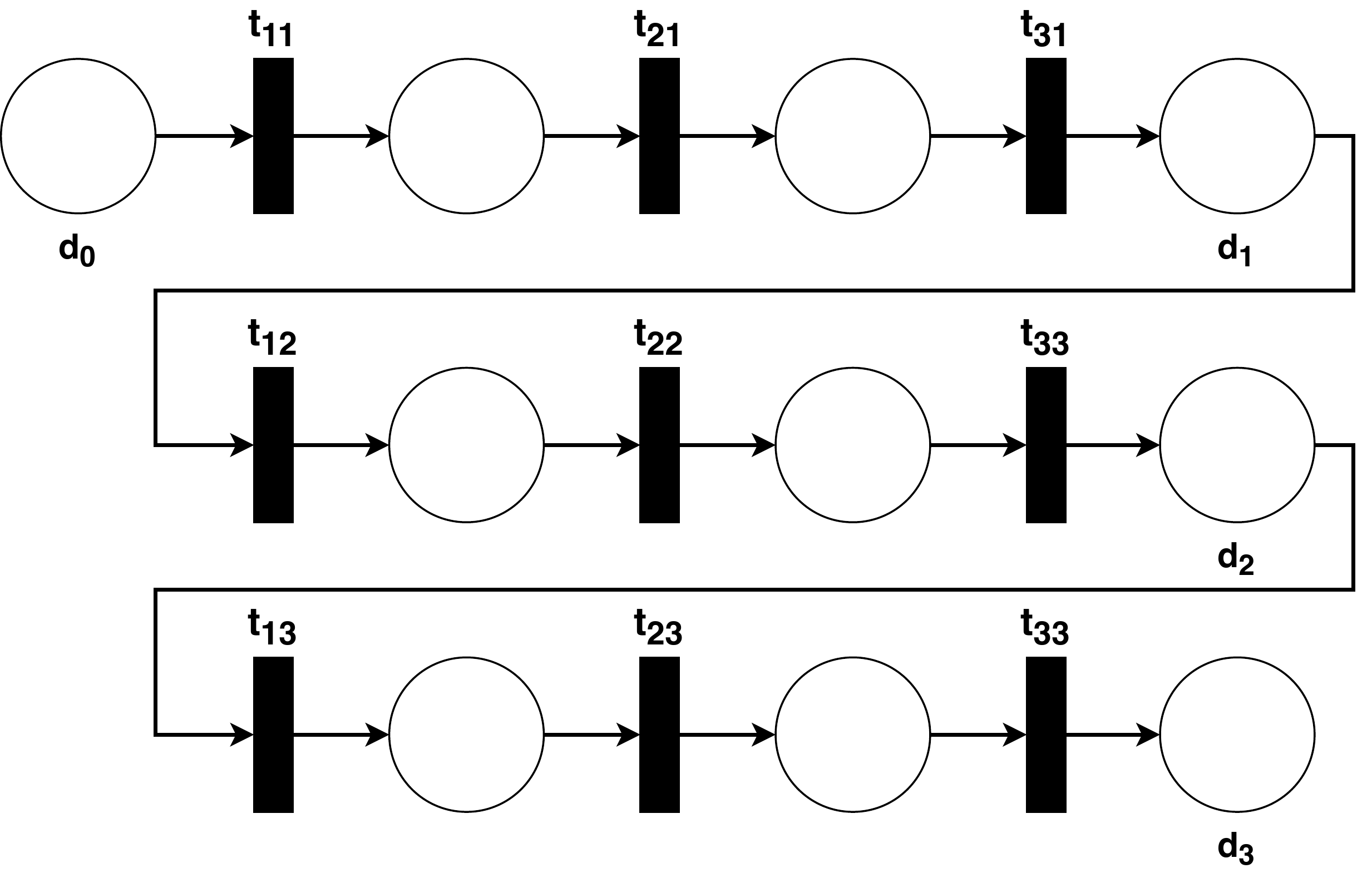}
\caption{Expanded iterative data flow for 3 iterations.}
\label{fig:expanded-iterate-petri}
\end{figure}

This model can be reduced into a model without cycles. This is true because all of the studied systems do require either a explicit limit of iterations ($n$) or, because the execution plan (which corresponds to the construction of the data flow model) is evaluated before the actual execution of the transformations. Consequently, the
execution plan always contains the information on the number of required iterations, making it possible to unfold the iteration as many times as needed.
For example, considering the iterative data flow shown in Figure~\ref{fig:iterate-petri}, when unfolding this data flow for 3 iterations ($n = 3$), we obtain the data flow shown in Figure~\ref{fig:expanded-iterate-petri}, in which the auxiliary transitions $t_{start}$, $t_{iterative}$ and $t_{end}$ were removed and the transformations within the step function have were repeated 3 times.

\paragraph{Iterative Transformations}
we define the semantics of iterative transformations in terms of the $\text{\textbf{repeat}}$ operation of monoid algebra, which receives a step function $f$ of type $Bag[\alpha] \rightarrow Bag[\alpha]$, a predicate function $p$ of type $Bag[\alpha] \rightarrow boolean$, a counter n ($n \in \mathbb{Z}$) and a bag $D$ of type $Bag[\alpha]$ as input and recursively applies the function $f$ until the condition in $p$ is reached or $n$ iterations occur, returning the resulting collection as output.

We define two iterative transformations: \textit{iterate} and \textit{iterateWithCondition}. The \textit{iterate} transformation takes a step function $st$, a counter $n$, and a collection $D$ as input and applies $st$ $n$ times. The  transformation \textit{iterateWithCondition} is similar, but it receives an additional predicate function $p$, so it iterates $n$ times or until the condition in $p$ is false, whichever is reached first ($n$ is necessary to avoid an infinite loop if $p$ is never reached). The definitions of \textit{iterate} and \textit{iterateWithCondition} are as follows:
\begin{align*}
\begin{split}
iterate ::{}& (Bag[\alpha] \rightarrow Bag[\alpha]) \rightarrow \mathbb{Z} \rightarrow Bag[\alpha] \rightarrow Bag[\alpha] \\
iterate(st, n, D) ={}& \text{\textbf{repeat}}(st, \lambda x. true, n, D)
\end{split}
\end{align*}

\begin{align*}
\begin{split}
iterateWithCondition ::{}& (Bag[\alpha] \rightarrow Bag[\alpha]) \rightarrow \\
{}& \,\,(Bag[\alpha] \rightarrow boolean) \rightarrow \\
{}& \,\,\,\mathbb{Z} \rightarrow Bag[\alpha] \rightarrow Bag[\alpha] \\
iterateWithCondition(st, p, n, D) ={}& \text{\textbf{repeat}}(st, p, n, D)
\end{split}
\end{align*}



\paragraph{Example}
to illustrate how an iterative program is represented in our model, let us consider the implementation of the \textit{PageRank} algorithm~\cite{brin1998anatomy} in Apache Spark presented in Figure~\ref{fig:pageRank-spark}. This version was based on the implementation presented in~\cite{Zaharia:2012}. The \textit{PageRank} algorithm calculates the importance (ranking) of a page based on the number of links from other pages to it. Rankings are calculated iteratively so that in each iteration, a page contributes to the ranking of the pages it links to and updates its ranking with the contribution it receives from the other pages that link to it. 

\begin{figure}[hbp!]
\centering
\begin{lstlisting}[style=Scala]
def pageRank(links: RDD[(String, Iterable[String])], n: Int) = {
    var ranks = links.map( link => (link._1, 1.0) )
    for(i <- 1 to n){
        val linksRanks = links.join(ranks)
        val values = linksRanks.map( lr => lr._2 )
        val contribs = values.flatMap { v =>
            val size = v._1.size
            v._1.map( url => (url, v._2 / size) )
        }
        val aggregContribs = contribs.reduceByKey( (a, b) => a + b )
        ranks = aggregContribs.map( rank => (rank._1, 0.15 + 0.85 * rank._2) )
    }
    ranks
}
\end{lstlisting}     
\caption{\textit{PageRank} implementation in Spark (based on~\cite{Zaharia:2012}).}
\label{fig:pageRank-spark}
\end{figure}

The program shown in Figure~\ref{fig:pageRank-spark} receives as input a key/value dataset of \textit{links}, where the key is the address of a page, and the value is the collection of pages it links to (line 1). The program also receives the number of iterations (\textit{n}) that will be made as input. The program starts by creating the initial \textit{ranks} dataset, in which each page (key) of the links dataset receives an initial ranking of $1.0$ (line 2). The iterative part is defined between lines 3 and 12, where the iterations are controlled through a \textit{for} statement executed from $1$ to \textit{n}. We abstract the block inside the \textit{for} statement (lines 4 to 11) as the step function that receives the \textit{ranks} dataset as input and produces, at the end of the iteration, a new version of the \textit{ranks} dataset with the updated ranking of each page as output. 

The step function starts with a join between \textit{links} and \textit{ranks} (line 4). 
Note that the dataset \textit{links} is not changed in the step function, but is only used in the \textit{join} with \textit{ranks}.
We have a dataset where each element is a tuple containing the page address, its ranking and the list of pages that it links to. Then we take only the part that contains the ranking and the list of links to other pages (line 5). After that, we calculate the contribution that each page sends to the ranking of the others pages it links to (line 6 to 9). This contribution is equal to $\frac{r}{s}$, where $r$ is the page ranking and $s$ is the number of neighbors (pages it links to). Next, we aggregate the contributions with the \textit{aggregateByKey} transformation (line 10). Since the \textit{contribs} dataset has key/value pairs where the key is a page and the value is the contribution it receives from another page, the result of the aggregation is a key/value dataset with the page (key) and the sum of all contributions it received (value). At the end of the step function (line 11), we update the \textit{ranks} dataset so that the ranking of each page is equal to $0.15 + 0.85 \times c$, where $c$ is the sum of all contributions received by the page. The program ends by returning the final \textit{ranks} dataset with the ranking of each page calculated after \textit{n} iterations (line 15).

To model the data flow of this program, we need to identify the datasets and transformations defined outside and inside the step function (iteration). Outside the step function, we have the input dataset $links$ of type $Bag[String \times Bag[String]]$ and the initial ranks dataset of type $Bag[String \times Double]$, defined before the iteration, which we call $ranks_0$. We also have the map transformation ($t_1$) that is applied to generate $ranks_0$. 

The datasets used within the step function ($st$) are updated by each iteration. 
Within the step function, we denote the datasets and transformations with an $i$ subscript, representing that at each iteration, a new version of the dataset or transformation will be created. 

In this example, we have the datasets $ranks_i: Bag[String \times Double]$ (note that is the same type of $ranks_0$), $linksRanks_i: Bag[String \times (Bag[String] \times Double)] $, $values_i: Bag[Bag[String] \times Double] $, $contribs_i: Bag[String \times Double]$ and $aggregContribs_i: Bag[String \times Double]$. 
In order to fit the iteration subnet pattern, we need to distinguish between the $ranks$ variable before iteration and after iteration. 
That gets us then the following set of places for our Petri Net:
\begin{align*}
D ={}& \{ links, ranks_0, ranks_{i-1}, linksRanks_i, values_i, contribs_i, \\
{}&   \,\,\,\,\,\, aggregContribs_i, ranks_i, ranks_n \}
\end{align*}

We also have the \textit{innerJoin} transformation $t_{2i}$, the \textit{map} $t_{3i}$, the \textit{flatMap} $t_{4i}$, the \textit{reduceByKey} $t_{5i}$ and the \textit{map} $t_{6i}$. 

\begin{figure}
\centering
\includegraphics[width=.9\textwidth]{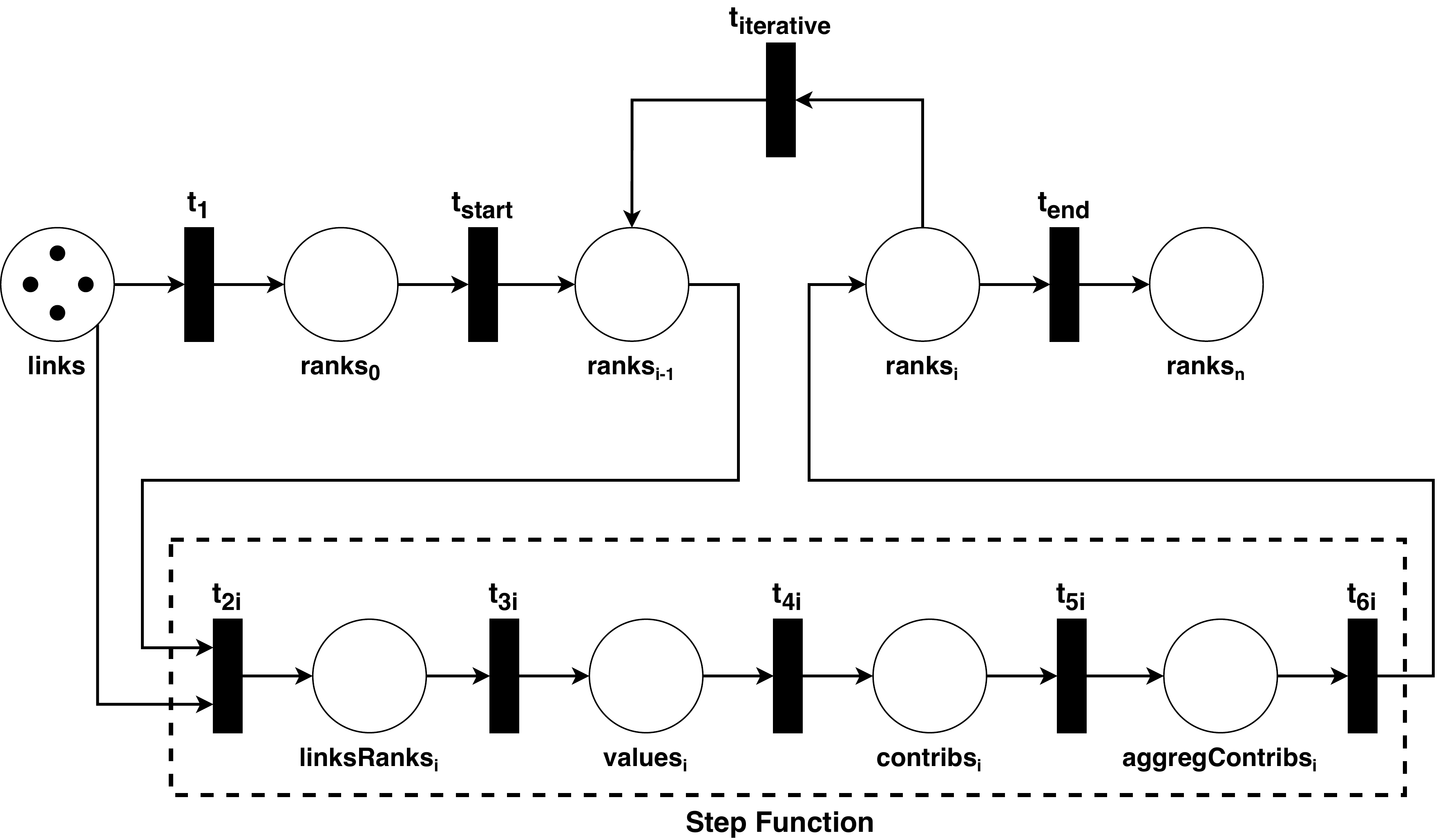}
\caption{Data flow of the PageRank program.}
\label{fig:petri-pageRank-spark}
\end{figure}

The data flow graph representing the \textit{PageRank} program is shown in Figure~\ref{fig:petri-pageRank-spark}. In it, we can see the data sets and transformations defined and the edges that connect them. We can also see the $t_{start}$, $t_{iterative}$ and $t_{end}$ transitions that represent the beginning, continuation and end of the iterations.
In terms of Monoid Algebra, the program is defined as follows:
\begin{align*}
t_1 ={}& map(\texttt{link => ...}, links) \\
t_{2i}(ranks_i) ={}& innerJoin(links, ranks_i) \\
t_{3i}(linksRanks_i) ={}& map(\texttt{lr => ...}, linksRanks_i) \\
t_{4i}(values_i) ={}& flatMap(\texttt{v => ...}, values_i) \\
t_{5i}(contribs_i) ={}& reduceByKey(\texttt{(a,b) => ...}, contribs_i) \\
t_{6i}(aggregContribs_i) ={}& map(\texttt{rank => ...}, aggregContribs_i) 
\end{align*}
where $i$ ranges from $1$ to $n$. 

\begin{figure}
\centering
\includegraphics[width=.9\textwidth]{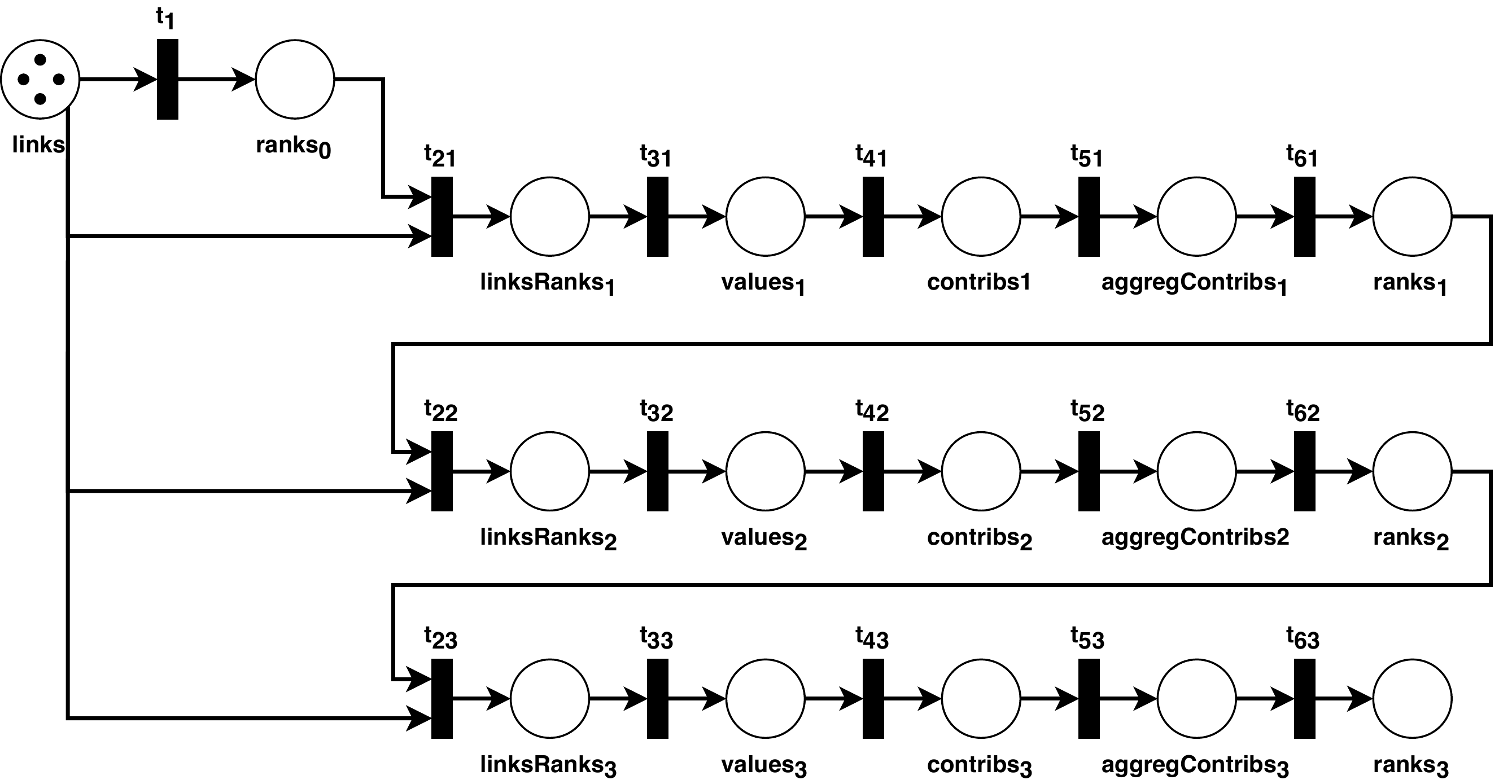}
\caption{Expanded data flow (without cycle) of the \textit{PageRank} program for 3 iterations.}
\label{fig:expanded-petri-pageRank-spark}
\end{figure}

The iteration that begins at $t_\textit{start}$ and ends at $t_\textit{end}$  is defined as:
\begin{align*}
t_{iterate} ={}& iterate(st, n, ranks_0), \text{ where } \\
st ={}& t_{6i} \circ t_{5i} \circ t_{4i} \circ t_{3i} \circ t_{2i}  
\end{align*}
As we mentioned earlier, the data flow systems that we are modeling define their programs as DAGs, so the representation of iterative programs takes place through the repetition of operations $n$ times where $n$ is the number of iterations, having no cycles in the graph as we did in our model. 
Our iteration representation is an abstraction for the expansion of the graph, but in fact, our model allows us to represent the DAG that would be created in the data flow systems. As an example, we can see the expanded representation of the data flow of the \textit{PageRanks} program for 3 iterations in Figure~\ref{fig:expanded-petri-pageRank-spark}. In it we can see that the iterations control transitions ($t_{start}$, $t_{iterative}$ and $t_{end}$) were removed and that the program is represented as a DAG.

The principles of the example given above are applicable to any structured iteration defined at the Petri Net level.
It is easy to see that the the transformation from an iterative Petri Net into an acyclic one, for a given $n$, can be defined using graph transformation/rewriting.


\section{Comparing Parallel Big Data Processing Frameworks}
\label{sec:bigdata}

The model proposed in this paper uses as reference the  characteristics of the programming strategies implemented by most prominent data flow based Big Data processing frameworks like \textit{Apache Spark}~\cite{Zaharia:2010}, \textit{Dryad/DryadLINQ}~\cite{Isard2007,yu2008}, \textit{Apache Flink}~\cite{katsifodimos2015apache} and \textit{Apache Beam}~\cite{beam2016}.  These frameworks use a similar DAG-based model to represent the data processing programs workflow despite the adoption of different strategies for executing programs, optimizing and processing data.  DAGs are composed of data processing operations that are connected through communication channels. The channels are places for intermediate data storage among operations.

Our model captures DAGs (data processing operations and communication channels) with the Petri Net data flow component. The nodes for datasets represent the communication channels  among operations. They represent at a high level, the abstractions used by Big Data processing frameworks for modeling distributed datasets, such as \textit{RDD} in \textit{Apache Spark} (see Figure~\ref{fig:union-logs-example} and Figure ~\ref{fig:union-logs-dataflow}),   \textit{PCollection} in \textit{Apache Beam}, \textit{DataSet} in \textit{Apache Flink} and \textit{DryadTable} in \textit{DryadLINQ}. Transformation nodes represent the processing operations that receive data from datasets and transmit the processing results to another dataset. The representations of the datasets and transformations in the data flow graph encompass the main abstractions of the DAGs in these systems and allow to represent and analyze a program independently of the system in which it will be executed. 
%
The semantics of transformations and data sets is represented in the model using Monoid Algebra. 

In this paper we focus on the abstract representation of both non-iterative and iterative Big Data processing programs.
Therefore, the following lines compare and discuss strategies adopted by existing frameworks for implementing this type of programs. They also discuss how our model provides a general formal specification of these strategies.

\subsection{Big Data Processing Frameworks}

Big Data processing frameworks adopt \textit{control flow} or \textit{data flow} based parallel programming models for implementing programs. Dependence analysis is a formal theory in compilation theory for determining ordering constraints between computations~\cite{kennedy2001optimizing}. The theory distinguishes between control and data dependencies. 
Control flow models focus on sequential (imperative) programming~\cite{Ivanovs2018}, thus the data follows the control and computations are executed explicitly based on the sequence programmed. 
Data flow models focus on data dependencies and allow avoiding spurious control dependencies like accidental locking~\cite{Ivanovs2018}, which simplifies the definition of concurrent and independent computations.  

\textit{Apache Hadoop}~\cite{hadoop273} is an open-source control flow system for the processing of distributed data that implements the \textit{MapReduce} programming model~\cite{Dean2004}.
MapReduce is a  parallel computing model that divides processing into two operations: \textit{map} and \textit{reduce}.
The \textit{map} operation applies the same function to all the elements of a list of key/value records.
The result of \textit{map} is feed to the \textit{reduce} operation which processes key/value data aggregated by the key. 
Other systems   such as Apache Spark~\cite{Zaharia:2010}, Apache Flink~\cite{katsifodimos2015apache}, Apache Beam~\cite{beam2016}, and Dryad/DryadLINQ~\cite{Isard2007, yu2008}, adopt data flow models that show  better performance.

Both control and data flow parallel programming models reach expression and execution limitations when implementing iterative data processing operations in many domains of data analysis, like machine learning or graph analysis. 
With increasing interest to run these kinds of algorithms on massive datasets, there is a need to execute iterations in a massively parallel fashion. Therefore, existing systems propose different strategies for implementing iterative operations. The following lines analyze and compare these strategies.


\paragraph{Apache Spark}
~\cite{Zaharia:2010} is a general purpose system for in-memory parallel data processing. 
Spark is centered on the concept of RDDs (\textit{Resilient Distributed Datasets}), which are distributed datasets that can be processed in parallel in a processing cluster. 
Spark programs are represented through a DAG that defines the program's data flow, where RDDs are processed by applying operations to them. 
Spark offers two types of operations.
\textit{Transformations}, which process the data in an RDD and generate a new RDD as output, and \textit{actions}, which save the contents of the RDD or generate a different result from an RDD.
Spark adopts a lazy evaluation strategy, where actions trigger the processing of data, possibly applying transformations.
For instance in the program given on Figure~\ref{fig:union-logs-example}, the program \textit{unionLogsProblem} defines three transformations (lines 2, 3 and 4) that will be executed when needed. This program encapsulates only the operations of transformations in RDDs. Its execution is triggered with the call of an action, which can be called later. An example is the \textit{collect} action that triggers the processing of transformations and collects the resulting RDD as a local data collection.



%

The in-memory processing of Spark proved to be more efficient than that of Apache Hadoop, making it more suitable for iterative programs since intermediate data does not need to be stored on disk~\cite{Zaharia:2010}, as occurs in Hadoop.  However, Spark does not have a native solution for defining iterative programs, making it necessary to use resources from the underlying programming language, like \textit{while} and \textit{for} loops, so that iterations can be defined. Since Spark adopts a lazy evaluation strategy, the definition of the data flow through the call of successive transformations forms an execution plan. This plan is optimized in a DAG and executed in parallel when an action is called. The definition of iterative programs follows the same principle. In this way, transformations called within an iteration form a step in the execution plan, making these transformations to be repeated in the DAG as many times as the number of iterations programmed in the loop (see the \textit{PageRank} example presented in Section~\ref{sec:ModelingIterations}).

\paragraph{Apache Beam}
~\cite{beam2016}  is a unified model for defining both batch and streaming data-parallel processing pipelines. Beam is  useful for implementing parallel data processing tasks, in which the problem can be decomposed into many smaller bundles of data that can be processed independently and in parallel. A pipeline can be executed by one of Beam's supported distributed processing back-ends, which include \textit{Apache Flink}, \textit{Apache Spark}, and \textit{Google Cloud Dataflow}.

Apache Beam programs are defined as data pipelines (Pipeline) that encapsulate its data flow with distributed data collections (PCollection) and data processing operations (PTransform). Thus, a program is defined by reading an input dataset, applying operations to datasets and writing an output dataset. This pipeline is optimized in a DAG and submitted for execution in a back-end engine.
Similar to Apache Spark, Beam does not provide a definitive solution for implementing iterative programs. Thus, the definition of iterative programs is based on the use of resources from the underlying programming language, relying on external control to the pipeline to control iterations.

\paragraph{Dryad/DryadLINQ}
~\cite{Isard2007} is a system and model for parallel and distributed programming that was proposed by \textit{Microsoft}. 
\textit{Dryad} offered a flexible programming model by representing a program through a DAG where the vertices are processing operations and the edges are communication channels through which data is transferred. 
With this model, a program is not limited to just two operations as in MapReduce. 
\textit{Dryad} was expanded through \textit{DryadLINQ}~\cite{yu2008}, a high-level interface that introduces an abstraction for representing distributed datasets (\textit{DryadTable}) and offered a comprehensive set of operations. 
A program in \textit{DryadLINQ} is represented by a data stream defined as a DAG, in which datasets are processed by applying operations in sequence.
The definition of iterative programs in Dryad/DryadLINQ also follows the approach of Apache Spark and Apache Beam, \textit{i.e.}, there is no native operation to control iterations, but they can be defined using loops from the underlying programming language.

\paragraph{Apache Flink}
is a framework and distributed processing engine for batch and streaming data processing~\cite{katsifodimos2015apache}. 
The system process arbitrary data flow programs in a distributed runtime environment.
As in other frameworks, the data flow is organized as a DAG with one or more entry or exit points.
Flink implements a lightweight fault tolerant model based on the use of \textit{checkpoints} that can be manually placed in the program or that can be added by the system.
Flink offers the DataSet API for batch processing and the DataStream API for streaming processing. Both offer a comprehensive set of operations for data processing, with mapping, filtering and aggregation operations, in addition to other types of operations.

From the Big Data processing frameworks analyzed in this work, Flink is the only one that offers a native solution for iterative programs.
For the definition of iterative programs, Flink offers the \textit{iterate} operation. This operation takes as an argument a high-order function, called step function, which encapsulates the iterative data flow that consumes an input dataset and produces an output dataset, which in turn is the input for the next iteration. The iterate operator implements a simple form of iterations: in each iteration, the step function consumes the entire input (the result of the previous iteration, or the initial dataset), and computes the next version of the partial solution. There are two options to specify termination conditions for an iteration specifying: (i) the maximum number of iterations, the iteration will be executed this many times; (ii) custom convergence function that implements a convergence criterion to end iterations.
Flink also offers the delta iterate operator (\textit{iterateDelta}) to address the case of incremental iterations that selectively modify elements of their solution and evolve the solution rather than fully recompute it. This leads to more efficient algorithms, because not every element in the solution set changes in each iteration.

Table~\ref{tab:systems-operations} compares the transformations defined by the model and the operations implemented in the Big Data processing frameworks.   Therefore, we grouped the transformations according to the types of processing that are done: \textit{Mapping}, \textit{Filtering}, \textit{Grouping}, \textit{Sets}, \textit{Aggregation}, \textit{Joins} and \textit{Ordering}. We modeled the main types of operations provided by these frameworks. In the table we also indicate how the model and frameworks deal with iterative programs.

\begin{table}
	\caption{Comparing   our model operations with  operations in Big Data processing frameworks.}
	\centering
	\begin{adjustbox}{width=\textwidth}
	\begin{tabular}{|c|p{2.5cm}|p{2.5cm}|p{2.5cm}|p{2.5cm}|p{2.5cm}|}
		\hline
		& \textbf{Model} & \textbf{Apache Spark} & \textbf{Apache Flink} & \textbf{Apache Beam} & \textbf{DryadLINQ} \\ \hline
		\textbf{Mapping} & map, flatMap & map, flatMap & map, flatMap & ParDo, FlatMapElements, MapElements & Select, SelectMany \\ \hline
		\textbf{Filtering} & filter & filter & filter & Filter & Where \\ \hline
		\textbf{Grouping} & groupBy, groupByKey & groupBy, groupByKey & groupBy & GroupByKey & GroupBy \\ \hline
		\textbf{Sets} & union, intersection, subtract, distinct & union, intersection, subtract, distinct & union, distinct & Flatten, Distinct & Union, Intersect, Except, Distinct \\ \hline
		\textbf{Aggregation} & reduce, redubeByKey & reduce, reduce\-ByKey, aggregateByKey & reduce, reduceGroup, aggregate & Combine & Aggregate \\ \hline
		\textbf{Joins} & innerJoin, leftOuterJoin, rightOuterJoin, fullOuterJoin & join, leftOuterJoin, rightOuterJoin, fullOuterJoin & join, leftOuterJoin, rightOuterJoin, fullOuterJoin & CoGroupByKey & Join \\ \hline
		\textbf{Ordering} & orderBy, orderByKey & sortBy, sortByKey & sortPartition, sortGroup &  & OrderBy \\	\hline
		\textbf{Iteration} & iterate, iterateWithCondition & Support with external for and while loops  & iterate, deltaIterate  & Support with external for and while loops & Support with external for and while loops \\ \hline 
	\end{tabular}
\end{adjustbox}
\label{tab:systems-operations}
\end{table}

Some systems offer more specific operations that we do not define directly in our model. It is a work in progress to guarantee complete coverage of all the operations of the considered systems. However, most of the operations that are not directly represented in the model can easily be represented using the transformations provided by the model. For example, classic aggregation operations, like maximum, minimum or the sum of the elements in a dataset. We can easily represent these operations using the \textit{reduce} operation of the model:
\begin{align*}
\begin{split}
max(D) ={}& reduce(\lambda (x, y) . \text{ \textbf{if} } x > y \text{ \textbf{then} } x  \text{ \textbf{else} } y, D) \\
min(D) ={}& reduce(\lambda (x, y) . \text{ \textbf{if} } x < y \text{ \textbf{then} } x  \text{ \textbf{else} } y, D) \\
sum(D) ={}& reduce(\lambda (x, y) . x + y, D)
\end{split}
\end{align*}

Ideally,  Big Data processing frameworks should allow users to express data flow using simple imperative data flow statements while matching the performance of native data flow. Therefore we believe that it is necessary to propose formal models agnostic of the underlying programming models and their implementation to reason about iterative and non-iterative data processing algorithms abstractly. The model proposed in the previous sections can be an abstraction of existing data flow-based programming models independently of their specific implementations by different frameworks. It provides abstractions of the data flow programming models that can be applied to specify parallel data processing programs independently of target systems.  

An abstract representation of parallel data flow based can be used for addressing program testing challenges beyond comparing Big Data processing tools that can be useful when adopting a framework and for migrating solutions from one framework to another.
Our model is used as a representation tool for defining mutation operators to apply mutation testing on data flow based Big Data processing programs.
In the next section, we briefly discuss how this is done in a testing tool we developed.

\section{Applications of the model}
\label{sec:exp}


The abstract and formal concepts provided by the model make it  suitable for the automation of software development processes, such as those done by IDE tools. 
Consequently,
we first applied the model to formalize the mutation operators presented in~\cite{caise2020}, where we explored the application of mutation testing in Spark programs, and in the tool \textsc{TRANSMUT-Spark}\footnote{\textsc{TRANSMUT-Spark} is publicly available at \url{https://github.com/jbsneto-ppgsc-ufrn/transmut-spark}.}~\cite{neto2020} that we developed to automate this process.
Mutation testing is a fault-based testing technique that relies on simulating faults to design and evaluate test sets~\cite{offut:2010}. Faults are simulated by applying mutation operators, which are rules with modification patterns for programs (a modified program is called a mutant). In~\cite{caise2020}, we presented a set of mutation operators designed for Spark programs that are divided into two groups: \textit{mutation operators for the data flow} and \textit{mutation operators for transformations}. These mutation operators were based on faults found in Spark programs with the idea of mimicking them.

\begin{figure}[!htbp]
	\centering
	\begin{subfigure}[t]{0.5\textwidth}
		\centering
		\includegraphics[width=.8\textwidth]{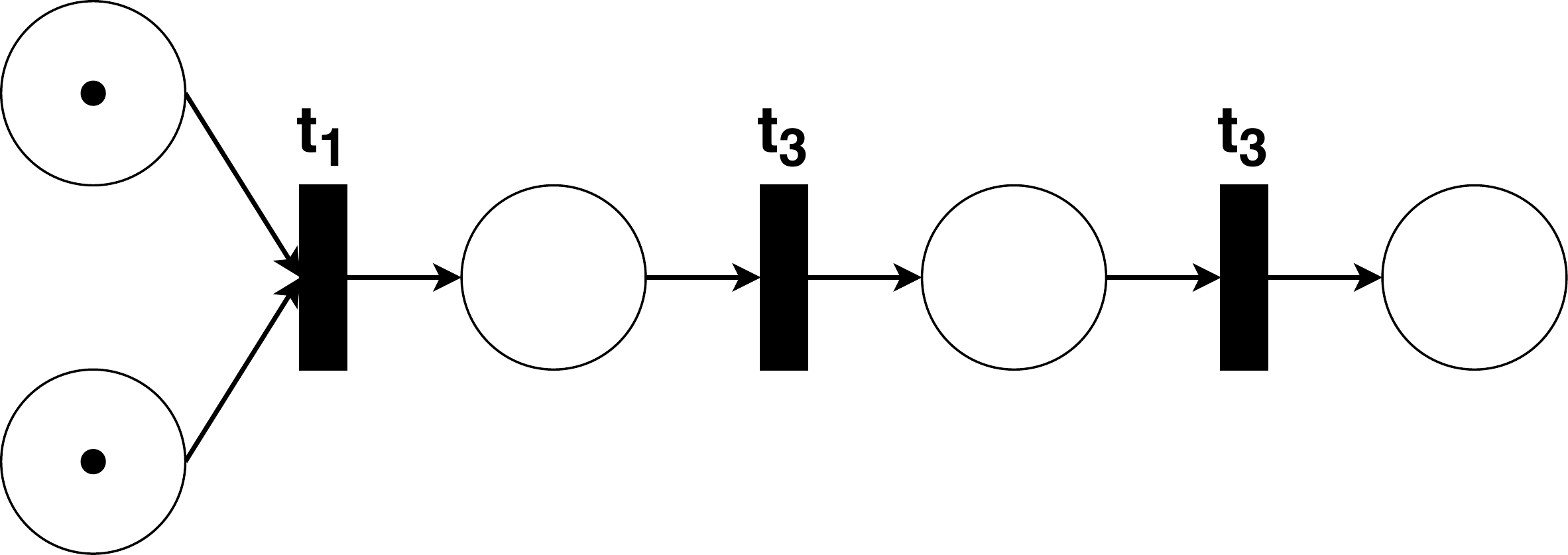}
		\caption{Transformation Replacement.}
		\label{subfig:dataflow-operators-1}
	\end{subfigure}%
	\hfill
	\begin{subfigure}[t]{0.5\textwidth}
		\centering
		\includegraphics[width=.8\textwidth]{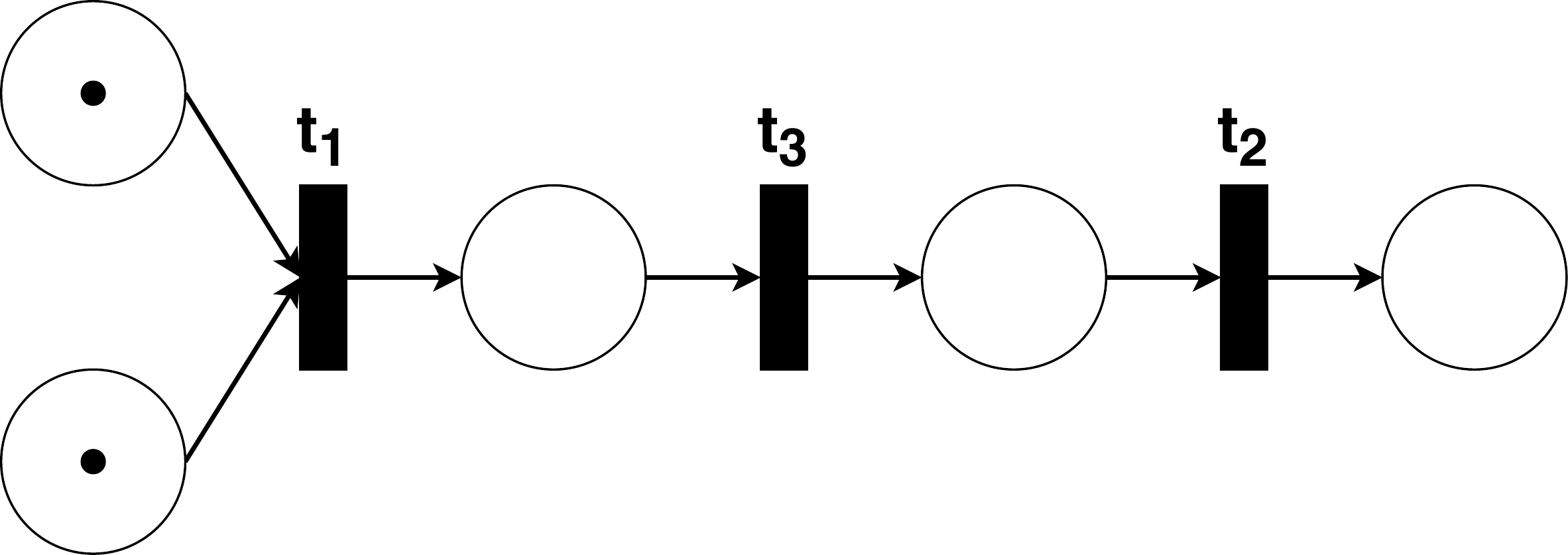}
		\caption{Transformations Swap.}
		\label{subfig:dataflow-operators-2}
	\end{subfigure}
	\\ 
	\begin{subfigure}[t]{0.5\textwidth}
		\centering
		\includegraphics[width=.6\textwidth]{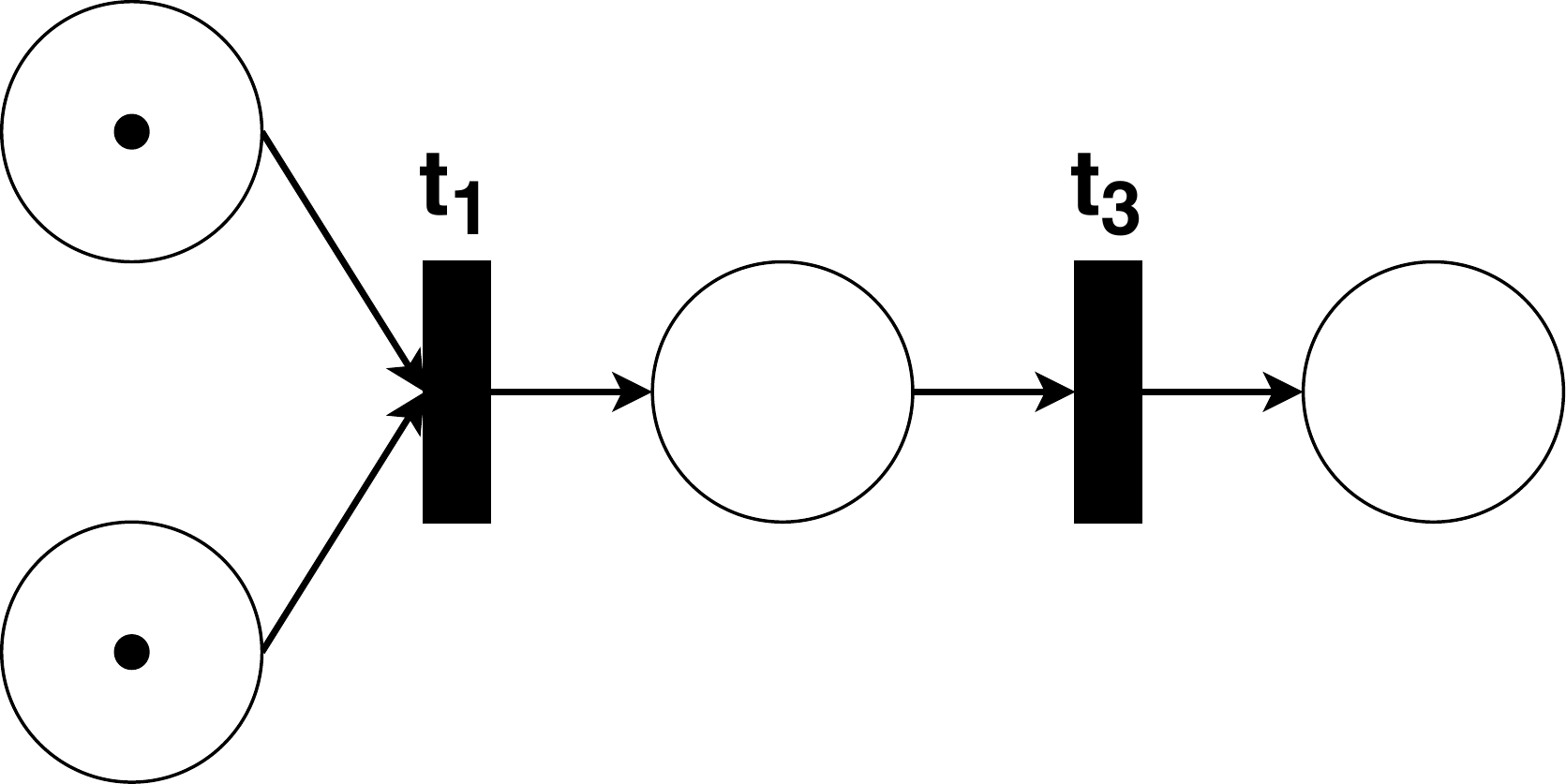}
		\caption{Transformation Deletion.}
		\label{subfig:dataflow-operators-3}
	\end{subfigure}
	\caption{Examples of mutants created with mutation operators for data flow.}%
	\label{fig:dataflow-operators}%
\end{figure}

Mutation operators for the data flow model change the DAG that defines the program.
In general, we define three types of modifications in the data flow: replacement of one transformation with another (both existing in the program), swap the calling order of two transformations and delete the call of a transformation in the data flow. These modifications involve changes to the edges of the program. Besides, the replacement of a transformation by another must maintain the type consistency, i.e., the I/O datasets of both transformations must be of the same type. In Figure~\ref{fig:dataflow-operators} we exemplify these mutations in the data flow that was presented in Figure~\ref{fig:union-logs-dataflow}.

Mutation operators associated with transformations model the changes done on specific transformations' types, such as operators for aggregation transformations or set transformations. In general, we model two types of modifications: replacement of the function passed as a parameter of the transformation and replacement of a transformation by another of the same group. In the first type, we defined specific substitution functions for each group of transformations. For example, for a transformation of type aggregation, we define five substitution functions ($f_m$) to replace it. Considering the aggregation transformation $t_1 = reduce(max(x, y), d)$, which receives as input a function that returns the greater of the two input parameters and an integer dataset, the mutation operator for aggregation transformation replacement will generate the following five mutants:
\begin{align*}
\begin{split}
t_1 ={}& reduce(f_m(x, y) = x, d) \\
t_1 ={}& reduce(f_m(x, y) = y, d) \\
t_1 ={}& reduce(f_m(x, y) = max(x, x), d) \\
t_1 ={}& reduce(f_m(x, y) = max(y, y), d) \\
t_1 ={}& reduce(f_m(x, y) = max(y, x), d)
\end{split}
\end{align*}

In the other type of modification, we replace a transformation with others from the same group. For example, for set transformations (union, intersection, and subtract), we replace one transformation with the remaining two; besides, we replace the transformation for the identity of each of the two input datasets, and we also invert the order of the input datasets. Considering the set transformation $t_1 = subtract(d_1, d_2)$, which receives two integer datasets a input, the set transformation replacement operator will generate the following mutants:
\begin{align*}
\begin{split}
t_1 ={}& union(d_1, d_2) \\
t_1 ={}& intersection(d_1, d_2) \\
t_1 ={}& identity(d_1) \\
t_1 ={}& identity(d_2) \\
t_1 ={}& subtract(d_2, d_1)
\end{split}
\end{align*}

The mutation operators for the other groups of transformations follow these two types of modifications, respecting each group's type consistency and particularities. The tool \textsc{TRANSMUT-Spark}~\cite{neto2020} uses the model as an intermediate representation. The tool reads a Spark program and translates it into an implementation of the model, so the mutation operators are applied to the model. We use the model as an intermediate representation in the tool to expand it in the future to apply the mutation test to programs in \textit{Apache Flink}, \textit{Apache Beam} and \textit{DryadLINQ}.

\section{Related Work}
\label{sec:relwork}

Data flow processing that defines a pipeline of operations or tasks applied on datasets, where tasks exchange data, has been traditionally formalized using (coloured) Petri Nets \cite{lee1987pipeline}.  They seem well adapted for modeling the organization (flow) of the processing tasks that receive and produce data. Regarding data processing programs based on data flow models, proposals use Petri Nets to model the flow and use other formal tools for modeling the operations applied on data. For example, \cite{hidders2005petri,hidders2008dfl} uses nested relational calculus for formalizing  operations applied to non first normal form compliant data. Next, we describe works that have formalized data processing parallel, programming models. The analysis focuses on the tools and strategies used for formalizing either control/data flows and data processing operations.

The authors in~\cite{yang2010} formalize MapReduce using \textit{CSP}~\cite{Hoare1984}. The objective is to formalize the behavior of a parallel system that implements the MapReduce programming model. The system is formalized with respect to four  components: \textit{Master}, \textit{Mapper}, \textit{Reducer} and \textit{FS} (file system). The Master manages the execution process and the interaction between the other components. The Mapper and Reducer components represent, respectively, the processes for executing the map and reduce operations. Finally, the FS represents the file system that stores the data processed in the program. These components implement the data processing pipeline implemented by these systems, loading data from an FS, executing a map function (by several mappers), shuffling and sorting, and executing a function reduce by reducers. The model allows the analysis of properties and interaction between these processes implemented by MapReduce systems.

In~\cite{Ono2011} MapReduce applications are formalized with \textit{Coq}, an interactive theorem proving systems. As in~\cite{yang2010}, the authors also formalized the components and execution process of  MapReduce systems. The user-defined functions of the map and reduce operations are also formalized with Coq. Then these formal definitions are used to prove the correctness of MapReduce programs. This approach is different from the work presented in~\cite{yang2010} (described above) that formalizes only the MapReduce system.

More recent work has proposed formal models for data flow programming models, particularly associated with Spark. The work in~\cite{Chen2017} introduces \textit{PureSpark}, a functional and executable specification for Apache Spark written in Haskell. The purpose of PureSpark is to specify parallel aggregation operations of Spark. Based on this specification, necessary and sufficient conditions are extracted to verify whether the outputs of aggregations in a Spark program are deterministic.

The work~\cite{Marconi2018} presents a formal model for Spark applications based on temporal logic. The model considers the DAG that forms the program, information about the execution environment, such as the number of CPU cores available, the number of tasks of the program and the average execution time of the tasks. Then, the model is used to check time constraints and make predictions about the program's execution time.


The research community has paid
attention to the  problem of addressing iterative programs in data flow based programming frameworks and have proposed a number of
solutions~\cite{alexandrov2019representations,moldovan2018autograph,jeong2019janus}. For example, Emma \cite{alexandrov2019representations} can translate
imperative control flow to Flink's native iterations, but only
when there is a single while-loop without any other control flow
statement in its body. This makes it not suitable for  
 data analytics tasks, such as hyper-parameter optimization,
simulated annealing, and strongly connected components.
AutoGraph \cite{moldovan2018autograph} and Janus \cite{jeong2019janus} compile imperative control
flow to TensorFlow’s native iterations \cite{yu2018dynamic}. However, they do
not support general data analytics other than machine learning. Mitos \cite{gevay2021efficient} allows users to write
imperative control flow constructs, such as regular while-loops and if statements.

\section{Conclusions and Future Work}
\label{sec:conc}

This paper presents a model for data flow processing programs.
Our model combines two formal mathematical tools: Monoid Algebra and Petri Nets.
Monoid Algebra is an abstract way to specify operations over partitioned datasets.
Petri nets are widely used to specify parallel computation. 
Our proposal combines these to models by building two-level specifications.
The lower level uses Monoid Algebra to specify individual transformations (\textit{i.e.}, operations whose arguments and results are datasets).
The upper level defines the program by means of a Petri Net, where places are datasets and transitions represent operations over that data.

The paper is an extended version of~\cite{deSouzaNeto2020sbmf}.
The main technical difference to that paper is the addition of iterations to the model, as well as the proposed use of our model to specify the operations available in several existing Big Data processing frameworks. In this sense, the paper gives the specification of data processing operations (i.e., transformations) provided as built-in operations  in Apache Spark, DryadLINQ, Apache Beam and Apache Flink.

In the proposed model, iterations are represented by a loop on the Petri Net that defines the program. Loops are \textit{unfolded} to build a Petri Net without cycles, to have a DAG representing the program.
This technique is convenient and realistic. It is convenient since it preserves the distribution and associative properties of the operations over datasets.
It is realistic since it provides a general model of strategies used by most prominent Big Data processing frameworks to process loops~\cite{Zaharia:2012,katsifodimos2015apache,beam2016,yu2008}.

\bigskip

Beyond the interest of providing a formal model for data flow-based programs, our proposal can be used as a comparison tool of target systems or to define program testing pipelines. 
We also showed how operations could be combined into data flows for implementing data mutation operations in mutation testing approaches.
The model was already used as an intermediary representation to specify mutation operators that were then implemented in \textsc{TRANSMUT-Spark}, a software engineering tool for mutation testing of Spark programs~\cite{caise2020}. Of course, the extension of the model to include iterations will also lead us to the definition of new mutation operators to cover the testing of iterative programs. 
Also, a natural extension to this work would be to instantiate the tool for other systems of the data flow family (\textit{DryadLINQ}, \textit{Apache Beam}, \textit{Apache Flink}).
This instantiation can be done by adapting \textsc{TRANSMUT-Spark}'s front and back ends so that a program originally written in any of these systems can be tested using the mutation testing approach proposed in~\cite{caise2020}. 
This line of work, where the model is used as the internal format, is suited for the more practical users, not willing to see the formalism behind their tools.  
However, when exploring the similarities of different frameworks, our model may be used as a platform-agnostic form of formally specifying and analyzing the properties of a program before its implementation.

\bigskip

As future work, we intend to study the extension of our model to use Colored Petri Nets (CPN) and CPN Tools~\cite{CPNTools} to specify the types for transformations over datasets explicitly and to manipulate, analyze and animate the specifications.
This extension may be useful to detect design problems at an early stage.
Also, we plan to work on the use of specifications for code generation to target data flow systems similar to Apache Spark.
A simple form of this code generation was implemented to generate test programs in \textsc{TRANSMUT-Spark} back-end~\cite{caise2020}.

 \bibliography{references}

\end{document}